\documentclass[notitlepage,nofootinbib,preprintnumbers,amssymb,superscriptaddress,eqsecnum]{revtex4-1}
\usepackage{amsfonts,amssymb,mathtools,graphicx,color,bm,autobreak,array}
\definecolor{ultramarine}{rgb}{0.07, 0.04, 0.56}
\definecolor{cadmiumgreen}{rgb}{0.0, 0.42, 0.24}
\definecolor{indigo(dye)}{rgb}{0.0, 0.25, 0.42}
\usepackage[linktocpage=true]{hyperref}
\hypersetup{
colorlinks=true,
citecolor=ultramarine,
linkcolor=cadmiumgreen,
urlcolor=indigo(dye),
}

\newcommand{\fr}[2]{\frac{#1}{#2}}
\newcommand{\D}{{\rm{d}}}
\newcommand{\pa}{\partial}
\newcommand{\ti}{\tilde}
\newcommand{\na}{\nabla}
\newcommand{\bra}[1]{\left( #1 \right)}  
\newcommand{\brb}[1]{\left[ #1 \right]}  
\newcommand{\brc}[1]{\left\{ #1 \right\}}  
\newcommand{\be}{\begin{equation}}  
\newcommand{\ee}{\end{equation}}
\newcommand{\bem}{\begin{bmatrix}}
\newcommand{\eem}{\end{bmatrix}}
\newcommand{\Mpl}{M_{\rm Pl}}

\newcommand{\ga}{\gamma}
\newcommand{\ep}{\epsilon}

\newcommand{\ka}{\kappa}
\newcommand{\la}{\lambda}
\newcommand{\si}{\sigma}

\newcommand{\Up}{\Upsilon}

\newcommand{\mn}{{\mu \nu}}
\newcommand{\mA}{\mathcal{A}}
\newcommand{\mB}{\mathcal{B}}

\newcommand{\mE}{\mathcal{E}}
\newcommand{\mF}{\mathcal{F}}
\newcommand{\mG}{\mathcal{G}}
\newcommand{\mH}{\mathcal{H}}

\newcommand{\mK}{\mathcal{K}}
\newcommand{\mL}{\mathcal{L}}
\newcommand{\mM}{\mathcal{M}}
\newcommand{\mO}{\mathcal{O}}

\newcommand{\mU}{\mathcal{U}}
\newcommand{\mV}{\mathcal{V}}
\newcommand{\mW}{\mathcal{W}}

\newcommand{\ve}[1]{{\rm{\bf{#1}}}}
\newcommand{\vu}{\ve{u}}
\newcommand{\bA}{\mathbb{A}}
\newcommand{\bB}{\mathbb{B}}
\newcommand{\bC}{\mathbb{C}}
\newcommand{\bI}{\mathbb{I}}
\newcommand{\bL}{\mathbb{L}}
\newcommand{\bP}{\mathbb{P}}
\newcommand{\bQ}{\mathbb{Q}}
\newcommand{\bR}{\mathbb{R}}
\newcommand{\PB}[1]{\brc{#1}_{\rm P}}

\begin{document}

\preprint{YITP-21-38}

\title{
\texorpdfstring{Black hole perturbations in DHOST theories: \\
Master variables, gradient instability, and strong coupling}
{Black hole perturbations in DHOST theories: 
Master variables, gradient instability, and strong coupling}
}

\author{Kazufumi Takahashi}
\affiliation{Center for Gravitational Physics, Yukawa Institute for Theoretical Physics, Kyoto University, Kyoto 606-8502, Japan}

\author{Hayato Motohashi}
\affiliation{Division of Liberal Arts, Kogakuin University, 2665-1 Nakano-machi, Hachioji, Tokyo 192-0015, Japan}

\begin{abstract}
We study linear perturbations about static and spherically symmetric black hole solutions with stealth scalar hair in degenerate higher-order scalar-tensor~(DHOST) theories.
We clarify master variables and derive the quadratic Lagrangian for both odd- and even-parity perturbations.
It is shown that the even modes are in general plagued by gradient instabilities, or otherwise the perturbations would be strongly coupled.
Several possible ways out are also discussed.
\end{abstract}

\maketitle

\section{Introduction}\label{sec:intro}

Starting from the first direct detection of gravitational waves from a binary black hole~(BH) merger~\cite{Abbott:2016blz}, so far a number of gravitational wave events have been observed.
Moreover, the first image of a BH was also obtained recently~\cite{Akiyama:2019cqa}.
These observations paved the way to test gravity at strong-field/dynamical regimes, which motivates us to study alternative theories of gravity and see how the modification affects the observables.
Although no deviation from exact solutions in general relativity~(GR) has been found so far, this does not mean all modified gravity theories are excluded.
This is because modified gravity theories in general allow a solution similar to or even of the same form as in GR.
In general, a solution with the metric of the same form as in GR is called ``stealth'' as the effect of modified gravity does not show up so long as one focuses on, e.g., the propagation of a test field in the background described by such a solution.
However, the behavior of gravitational perturbations about a stealth solution is in principle different from the one in GR. 
Namely, features of underlying gravitational theories are encoded in perturbations.
Hence, it is important to clarify whether the underlying theory is GR or not with precise observational data promised in the near future.

In order to study stealth solutions, we first need to specify the theory space to be surveyed.
According to the Lovelock theorem~\cite{Lovelock:1971yv}, GR plus a cosmological constant is the only metric theory with general covariance and having  second-order Euler-Lagrange~(EL) equations in four-dimensional spacetime.
Note that the second-order nature of EL equations is desirable as theories with higher derivatives in equations of motion~(EOMs) in general lead to the problem of Ostrogradsky ghost,\footnote{Note that a recent generalization~\cite{Aoki:2020gfv} of the Ostrogradsky theorem applies not only to ghost associated with higher-order derivatives, but also to ghost originating from lower-order derivatives in systems with constraints.}
which makes the Hamiltonian unbounded both at the classical~\cite{Woodard:2015zca} and quantum levels~\cite{Motohashi:2020psc}.
Therefore, to go beyond GR, one has to relax any of the assumptions of the Lovelock's theorem.
Interestingly, modified gravity theories in general, at least effectively, contain one or more additional degrees of freedom~(DOFs) on top of the metric.
For instance, theories without general covariance can be covariantized by introducing St{\"u}ckelberg field(s), while a dimensional reduction of a higher-dimensional theory yields an additional field corresponding to the volume of the extra dimension(s).
Hence, it would be useful to study scalar-tensor theories~(i.e., those with a single scalar field in addition to the metric) to capture aspects of various modified gravity theories.
In order for a scalar-tensor theory to be free of Ostrogradsky ghost, higher-derivative terms in the action must be degenerate~\cite{Motohashi:2014opa,Langlois:2015cwa,Motohashi:2016ftl,Klein:2016aiq,Motohashi:2017eya,Motohashi:2018pxg}.
Note that, even if the theory is viewed as an effective field theory~(EFT) valid up to some finite energy scale, the degeneracy conditions should be maintained below the cutoff scale.
Applying the degeneracy conditions to a given action, one obtains degenerate higher-order scalar-tensor~(DHOST) theories~\cite{Langlois:2015cwa,Crisostomi:2016czh,BenAchour:2016fzp,Takahashi:2017pje,Langlois:2018jdg}, which provide a general class of healthy scalar-tensor theories with higher derivatives.\footnote{One could still extend the framework by requiring the degeneracy only in the unitary gauge where the scalar field is a function only of time~\cite{Gao:2014soa,DeFelice:2018mkq,Gao:2018znj,Motohashi:2020wxj}.
In that case, there is an apparent extra DOF in a generic gauge, but it satisfies an elliptic differential equation and hence is an instantaneous mode.
In the present manuscript, we do not consider such a possibility since there is a subtlety in the treatment of instantaneous modes.}
Indeed, it includes Horndeski~\cite{Horndeski:1974wa,Deffayet:2011gz,Kobayashi:2011nu} and Gleyzes-Langlois-Piazza-Vernizzi~(GLPV) theories~\cite{Gleyzes:2014dya} as special subclasses (see \cite{Langlois:2018dxi,Kobayashi:2019hrl} for reviews).

Stealth solutions in scalar-tensor theories have been extensively investigated so far.
A trivial class is those accompanied by a constant scalar field, for which a set of existence conditions was derived in \cite{Motohashi:2018wdq}.
It is also possible to construct stealth solutions supporting nontrivial scalar hair.
The simplest one would be those with a constant kinetic term of the scalar field.
For instance, stealth BH solutions of this type have been obtained for the Schwarzschild--(anti-)de~Sitter spacetime~\cite{Babichev:2013cya,Kobayashi:2014eva,Babichev:2016kdt,Babichev:2017guv,Babichev:2017lmw,BenAchour:2018dap,Motohashi:2019sen,Minamitsuji:2019shy,Khoury:2020aya} and for the Kerr-de~Sitter spacetime~\cite{Babichev:2017guv,Charmousis:2019vnf} as well.
Moreover, existence conditions for stealth solutions in the presence of a general matter component were clarified in \cite{Takahashi:2020hso}.
Perturbations around such stealth solutions have been studied in \cite{Ogawa:2015pea,Takahashi:2015pad,Takahashi:2016dnv,Tretyakova:2017lyg,Babichev:2017lmw,Babichev:2018uiw,Minamitsuji:2018vuw,Takahashi:2019oxz,deRham:2019gha,Charmousis:2019fre,Motohashi:2019ymr,Khoury:2020aya,Tomikawa:2021pca}.
Although the full analysis for even-parity perturbations about stealth Schwarzschild--(anti-)de Sitter solutions is still lacking, it was found that the perturbations for the scalar mode have a vanishing sound speed at least for some class of theories, which indicates the problem of strong coupling~\cite{Babichev:2018uiw,Minamitsuji:2018vuw,deRham:2019gha}.
The origin of this strong coupling issue for stealth solutions was elucidated in \cite{Motohashi:2019ymr} by studying the perturbation dynamics in the asymptotic region.
Namely, if the underlying theory forces the perturbations to follow second-order EOMs, which is the case for theories satisfying the degeneracy conditions, either the strong coupling or gradient instability is inevitable in the asymptotic flat or de~Sitter region, and one cannot trust the stealth solutions within the regime of validity of the EFT.
Also, the same result for perturbations about the stealth Minkowski solution was obtained by constructing an EFT from the perspective of spacetime symmetry breaking~\cite{ArkaniHamed:2003uy}.
Therefore, to avoid the strong coupling/gradient instability, one needs to introduce a detuning term dubbed the ``scordatura'' term~\cite{Motohashi:2019ymr}, which weakly violates the degeneracy conditions in a controlled manner.
From the EFT viewpoint, it is natural to expect such a detuning unless the degeneracy conditions are protected by some fundamental symmetry.
The scordatura term modifies the dispersion relation and hence renders the strong coupling scale sufficiently high, yet maintaining the ghost-free nature below the EFT cutoff scale.\footnote{In the context of the stealth cosmological solution, the scordatura term is also necessary to make the quasi-static limit well-defined, implying that the subhorizon observables are inevitably affected by the scordatura~\cite{Gorji:2020bfl}.}
However, before introducing the scordatura term, it is necessary to perform the full analysis of perturbations about stealth BHs in DHOST theories.
In particular, for the even-parity perturbations about stealth Schwarzschild--(anti-)de Sitter solutions, the master variables and the stability have been veiled.\footnote{See \cite{Langlois:2021xzq,Langlois:2021aji} for a complimentary approach, by which 
the asymptotic behavior of perturbations can be extracted without specifying master variables.}

The aim of the present paper is to complete the perturbation analysis for stealth BH solutions in DHOST theories.
As clarified in \cite{Takahashi:2019oxz}, under several ansatz on the metric and scalar field profile, the shift- and reflection-symmetric subclass of quadratic DHOST theories admits the stealth Schwarzschild--(anti-)de~Sitter solution only.
We focus on such a unique stealth BH solution as the background solution, and establish the perturbation theory for both odd- and even-parity perturbations about the stealth BHs in DHOST theories by clarifying three master variables, one for the odd modes and the other two for the even modes.
Moreover, we show that the even modes are generically plagued by gradient instabilities, or otherwise the sound speed for scalar waves is tiny everywhere.
In the latter case, the perturbations are presumably strongly coupled, and hence the scordatura effect should be taken into account.

The rest of this paper is organized as follows.
In \S\ref{sec:BG}, we specify the model and briefly review the stealth BH solutions found in \cite{Takahashi:2019oxz}.
Then, in \S\ref{sec:odd} and \S\ref{sec:even}, we study odd- and even-parity perturbations about the stealth BHs, respectively.
We derive the quadratic Lagrangians in terms of master variables and demonstrate that the even modes in general suffer from gradient instabilities, or otherwise the perturbations would be strongly coupled.
To complete the analysis of the even-parity perturbations, we employ several techniques, which are summarized in the Appendices~\ref{AppA}--\ref{AppC}.
In Appendix~\ref{AppA}, we provide the transformation law of perturbation variables under the conformal/disformal transformation, which helps us to reduce the complicated quadratic Lagrangian for BH perturbations in DHOST theories.
In Appendix~\ref{AppB}, we perform a Hamiltonian analysis for even-parity perturbations to clarify the number of physical DOFs.
In Appendix~\ref{AppC}, we apply the method of characteristic analysis to monopole perturbations.
Finally, we draw our conclusions in \S\ref{sec:conc}.

\section{Stealth black holes in quadratic DHOST theories}\label{sec:BG}

\subsection{The model}\label{ssec:model}

We consider a subclass of quadratic DHOST theories~\cite{Langlois:2015cwa}, whose action has the following form:
	\be
	S=\int \D^4x\sqrt{-g}\brb{F_0(X)+F_2(X)R+\sum_{I=1}^{5}A_I(X)L_I^{(2)}}, \label{qDHOST}
	\ee
where $F_0$, $F_2$, and $A_I$ ($I=1,\cdots,5$) are functions of $X\coloneqq\phi_\mu\phi^\mu$ and 
	\be
	L_1^{(2)}\coloneqq \phi^{\mn}\phi_{\mn},~~~L_2^{(2)}\coloneqq (\Box\phi)^2,~~~L_3^{(2)}\coloneqq \phi^\mu\phi_{\mn}\phi^\nu\Box\phi,~~~L_4^{(2)}\coloneqq \phi^\mu\phi_{\mn}\phi^{\nu\la}\phi_\la,~~~L_5^{(2)}\coloneqq (\phi^\mu\phi_{\mn}\phi^\nu)^2,
	\ee 
with $\phi_\mu\coloneqq \na_\mu\phi$ and $\phi_{\mn}\coloneqq \na_\mu\na_\nu\phi$.
Clearly, this action is invariant under the shift (i.e., $\phi\to \phi+{\rm const}$) and the reflection (i.e., $\phi\to -\phi$) of the scalar field.
Among the coefficient functions in \eqref{qDHOST}, $A_2$, $A_4$, and $A_5$ are written in terms of the others as
	\be
	\begin{split}
	A_2&=-A_1\ne -\fr{F_2}{X}, \\
	A_4&=\fr{1}{8(F_2-XA_1)^2}\bigl\{4F_2\brb{3(A_1-2F_{2X})^2-2A_3F_2}-A_3X^2(16A_1F_{2X}+A_3F_2) \\
	&~~~~~~~~~~~~~~~~~~~~~~~~+4X\bra{3A_1A_3F_2+16A_1^2F_{2X}-16A_1F_{2X}^2-4A_1^3+2A_3F_2F_{2X}}\bigr\}, \\
	A_5&=\fr{1}{8(F_2-XA_1)^2}(2A_1-XA_3-4F_{2X})\brb{A_1(2A_1+3XA_3-4F_{2X})-4A_3F_2},
	\end{split} \label{DC}
	\ee
where a subscript~$X$ denotes a derivative with respect to $X$.
These conditions, which are referred to as degeneracy conditions, ensure the absence of Ostrogradsky ghost.
As a result, the model is characterized by four arbitrary function~$F_0$, $F_2$, $A_1$, and $A_3$.
The condition~$A_1\ne F_2/X$ (or equivalently, $F_2-XA_1\ne 0$) is necessary for the existence of two tensor modes on a cosmological background~\cite{deRham:2016wji}.
The class of theories described by \eqref{qDHOST} and satisfying \eqref{DC} is known as ``class Ia''~\cite{Achour:2016rkg} (also called ``class ${}^2$N-I''~\cite{Crisostomi:2016czh,BenAchour:2016fzp}).
This class includes the Horndeski theory~\cite{Horndeski:1974wa,Deffayet:2011gz,Kobayashi:2011nu} up to quadratic-order interactions. 
Indeed, the action reduces to that of Horndeski theory by imposing $A_1=2F_{2X}$ and $A_3=0$.
There are still other classes of quadratic DHOST theories~\cite{Langlois:2015cwa}, but they are phenomenologically undesirable since they are plagued by ghost/gradient instabilities on a cosmological background or otherwise have no propagating tensor DOFs~\cite{Langlois:2017mxy,deRham:2016wji}.
Hence, throughout the present paper, we focus on the class Ia of quadratic DHOST theories characterized by the degeneracy conditions~\eqref{DC}.

\subsection{Stealth black hole solutions}\label{ssec:BH}

In \cite{Takahashi:2019oxz}, the authors of the present paper studied general static spherically symmetric vacuum solutions in the quadratic DHOST theory.
For the background metric and scalar field of the form,
    \be
    g^{(0)}_{\mn}\D x^\mu \D x^\nu=
    -A(r)\D t^2+\fr{\D r^2}{B(r)}+r^2\ga_{ab}\D x^a\D x^b, \qquad
    \phi^{(0)}=qt+\psi(r), \label{ansatz}
    \ee
with $\gamma_{ab}$ being the metric on a two-dimensional sphere, we derived the EOMs and reduced them to a simple system of first-order differential equations for $A(r)$, $B(r)$, and $X(r)$.
From the reduced EOMs, we see that the following uniqueness theorem holds~\cite{Takahashi:2019oxz}: Under the ansatz~\eqref{ansatz} and the condition that $X$ is a negative constant and $F_2A_3+2(F_2A_1)_X\ne 0$, a generic DHOST theory described by the action~\eqref{qDHOST} admits the stealth Schwarzschild--(anti-)de~Sitter solution (including the asymptotically flat case) only as the static spherically symmetric vacuum solution, up to coordinate redefinition.
Specifically, the unique solution is given by 
    \be
    A(r)=B(r)=1-\fr{\mu}{r}-\fr{\Lambda}{3}r^2, \qquad
    X=-q^2, \label{stealthSdS}
    \ee
where $\mu$ is a positive constant, $q^2$ is a solution to the following algebraic equation:
	\be
	F_0(2A_1+4q^2A_{1X}+3q^2A_3-8F_{2X})+4F_{0X}(F_2+q^2A_1)=0, \label{exist}
	\ee
and the effective cosmological constant~$\Lambda$ is given by
    \be
    \Lambda=-\fr{F_0}{2(F_2+q^2A_1)}. \label{Lambda}
    \ee
Note that, for the case with $F_2A_3+2(F_2A_1)_X=0$, there exist Schwarzschild--(anti-)de~Sitter solutions with a deficit solid angle accompanied by constant $X$, where the parameter~$q$ remains arbitrary and the deviation of $X$ from $-q^2$ measures the amount of deficit.
Hence, the above uniqueness does not hold in this particular case.
Nevertheless, if we choose $q$ so that $X=-q^2$, the configuration~\eqref{stealthSdS} is recovered, and then the following discussion applies.
Therefore, in what follows, we do not exclude the case~$F_2A_3+2(F_2A_1)_X=0$ unless otherwise stated.

Given the metric, one can obtain $\psi(r)$ by integrating the condition $X=-q^2$.\footnote{In general, there are two branches for the solution of $\psi(r)$. We choose the plus branch so that the scalar field profile takes the form
\[ \phi^{(0)}=q\bra{t+\int \frac{\sqrt{1-A}}{A} \D r }. \] }
Here and hereafter, the coefficient functions of the DHOST theory (i.e., $F_0$, $F_2$, $A_1$, and $A_3$) and their derivatives are evaluated at $X=-q^2$ unless otherwise stated.
In the present paper, we focus on the stealth Schwarzschild-de~Sitter solution above and study perturbations around it.
For this purpose, it is useful to introduce the Lema\^{i}tre coordinates~\cite{Lemaitre:1933gd,Khoury:2020aya}, where the metric has the Gaussian normal form,
    \be
    g^{(0)}_{\mn}\D x^\mu \D x^\nu=
    -\D\tau^2+\brb{1-A(r)}\D\rho^2+r^2\ga_{ab}\D x^a\D x^b, \qquad
    \phi^{(0)}=q\tau. \label{Lemaitre}
    \ee
Here, the new set of coordinate variables~$(\tau,\rho)$ is related to $(t,r)$ through
    \be
    \D\tau=\D t+\fr{\sqrt{1-A(r)}}{A(r)}\D r, \qquad
    \D\rho=\D t+\fr{\D r}{A(r)\sqrt{1-A(r)}}.
    \ee
From this relation, we obtain
    \be
    \D(\rho-\tau)=\fr{\D r}{\sqrt{1-A(r)}},
    \ee
which implies that the original radial coordinate~$r$ is a function of $\rho-\tau$.
Note that
    \be
    r'=-\dot{r}=\sqrt{1-A(r)},
    \ee
where a dot (and a prime) denotes a derivative with respect to $\tau$ (and $\rho$, respectively).
We employ this notation throughout the present paper.
Therefore, for instance, $A'$ is not $\D A/\D r$ but
\be \label{Aprime} A'= \frac{\D A}{\D r} r'. \ee
Note also that we restrict ourselves to $\Lambda\ge 0$ since otherwise one cannot choose the Lema{\^i}tre coordinates globally as $1-A(r)$ can be 
negative for $\Lambda< 0$.
As we shall see in the next section, the stability of odd-parity perturbations requires $F_2+q^2A_1>0$, which from \eqref{Lambda} implies that $\Lambda\ge 0$ amounts to $F_0\le 0$.

\section{Odd-parity perturbations}\label{sec:odd}

We separate the deviation of the metric from its background, $h_{\mn}\coloneqq g_{\mn}-g^{(0)}_{\mn}$, into the odd- and even-parity modes
as they are completely decoupled from each other unless the Lagrangian contains parity-violating terms leading to analysis of coupled equations~\cite{Motohashi:2011pw,Motohashi:2011ds}. 
The perturbation analysis of the odd modes for general static spherically symmetric BHs has been established in \cite{Takahashi:2019oxz}.
Here, we redo the analysis for the particular case of stealth Schwarzschild-de~Sitter BHs in the Lema\^{i}tre coordinates.
As we shall see below, the analysis in the Lema\^{i}tre coordinates is more transparent and provides a weaker set of conditions for no ghost/gradient instabilities compared to the analysis in \cite{Takahashi:2019oxz} based on the static coordinates.

The odd-parity perturbations can be decomposed as follows:
	\be
	\begin{split}
	h_{\tau \tau }&=h_{\tau \rho }=h_{\rho \rho }=0,\\
	h_{\tau a}&=\sum_{\ell,m}r^2h_{0,\ell m}(\tau,\rho)E_a{}^b\bar{\na}_bY_{\ell m}(\theta,\varphi),\\
	h_{\rho a}&=\sum_{\ell,m}r^2h_{1,\ell m}(\tau,\rho)E_a{}^b\bar{\na}_bY_{\ell m}(\theta,\varphi),\\
	h_{ab}&=\sum_{\ell,m}r^2h_{2,\ell m}(\tau,\rho)E_{(a}{}^c\bar{\na}_{|c|}\bar{\na}_{b)}Y_{\ell m}(\theta,\varphi),
	\end{split} \label{pert_odd}
	\ee
where $Y_{\ell m}$ is the spherical harmonics, $E_{ab}$ is the completely antisymmetric tensor defined on a two-dimensional sphere, and $\bar{\na}_a$ denotes the covariant derivative with respect to $\gamma_{ab}$.
Here, in comparison to \cite{Takahashi:2019oxz}, we adopt a slightly different definition of $h_{0}$, $h_{1}$, and $h_{2}$ by factoring out $r^2$, which simplifies calculations in the Lema\^{i}tre coordinates [see, e.g., \eqref{gaugetrnsf_odd} and \eqref{qlag_odd}].
Since modes with different $(\ell,m)$ evolve independently, we focus on a specific mode and omit the indices~$\ell$ and $m$ unless necessary.
Note that the odd-parity perturbations do not have the monopole~($\ell=0$) mode and $h_2$ is vanishing for the dipole~($\ell=1$) modes.
Note also that the perturbation of the scalar field is absent as it belongs to the even-parity perturbations.

The expansion coefficients $h_0$, $h_1$, and $h_2$ are not all physical DOFs as there exists a gauge DOF associated with the general covariance.
The odd-parity part of an infinitesimal coordinate transformation~$x^\mu\to x^\mu+\ep^\mu$ can be written as
	\be
	\ep^\tau=\ep^\rho=0, \qquad
	\ep^a=\sum_{\ell,m}\Xi_{\ell m}(\tau,\rho)E^{ab}\bar{\na}_bY_{\ell m}(\theta,\varphi).
	\ee
Correspondingly, the gauge transformation of the coefficients $h_0$, $h_1$, and $h_2$ is given by
    \be
    h_0\rightarrow h_0-\dot{\Xi}, \qquad
	h_1\rightarrow h_1-\Xi', \qquad
	h_2\rightarrow h_2-2\Xi. \label{gaugetrnsf_odd}
    \ee

Here, we make a remark on our strategy for gauge fixing.
It is always valid to impose a gauge-fixing condition after deriving EL equations.
However, in this approach one needs to handle redundant EL equations containing gauge DOFs and hence the analysis tends to be cumbersome especially in theories of modified gravity. 
Hence, we would like to fix the gauge at the Lagrangian level before deriving EL equations and study only independent EL equations without gauge DOFs from the outset.
In general, we can classify gauge fixings into two classes: complete and incomplete gauge fixings.
The separation criterion between the two is whether the gauge-fixing condition fully determines the gauge functions or not.
This difference is inherited to the structure of the Noether identity for EL equations, from which we can deduce that 
a gauge fixing at the Lagrangian level is legitimate if it is complete, while illegitimate if it is incomplete since it causes a loss of independent EL equations~\cite{Motohashi:2016prk}.

Specifically, in the case of $\ell\ge 2$, one can fix the gauge by choosing $\Xi=h_2/2$ to set $h_2\to 0$.
This gauge fixing determines the gauge function $\Xi(\tau,\rho)$ without ambiguities. 
Thus, it is a complete gauge fixing and can be imposed at the Lagrangian level~\cite{Motohashi:2016prk}.
We shall use this gauge fixing for $\ell\ge 2$ in \S\ref{ssec:multipole_odd}.
On the other hand, for the dipole modes, since $h_2$ is absent we need to remove the gauge DOF by imposing a different condition.
From \eqref{gaugetrnsf_odd}, possible choices would be $\dot{\Xi}=h_0$ or $\Xi'=h_1$ to make $h_0\to 0$ or $h_1\to 0$, respectively.
However, in either case, $\Xi(\tau,\rho)$ is determined only up to a $\rho$- or $\tau$-dependent function as an integration constant.
Such a gauge fixing is incomplete, and should not be imposed at the Lagrangian level.
Therefore, for the dipole analysis in \S\ref{ssec:dipole_odd}, we fix the gauge after deriving EL equations.

\subsection{\texorpdfstring{Odd-parity perturbations with $\ell\ge 2$}{Multipole}}
\label{ssec:multipole_odd}

First, we consider higher multipoles with $\ell\ge2$.
One can set $m=0$ from the beginning since all the terms with the same multipole index~$\ell$ contributes equally by virtue of the spherical symmetry of the background.
Hence, it is more useful to expand the metric perturbations in terms of the Legendre polynomials instead of the spherical harmonics. 
Thus, in the subsequent analysis, $h_0$, $h_1$, and $h_2$ denote the coefficients of $P_\ell(\cos \theta)$.
After performing the integration over the angular variables, the quadratic action for odd modes takes the form
	\be
	S^{(2)}_{\rm odd}=\int \D\tau \D\rho \, \mL^{(2)}_{\rm odd},
	\ee
where
	\be
	\fr{2\ell+1}{2\pi j^2}\mL^{(2)}_{\rm odd}=
	p_1h_0^2+p_2h_1^2+p_3\bra{\dot{h}_1-h'_0}^2, \label{qlag_odd}
	\ee
with $j^2\coloneqq \ell(\ell+1)$ and the coefficients~$p_1$, $p_2$, and $p_3$ given by
    \be
    p_1=(j^2-2)(F_2+q^2A_1)r^2\sqrt{1-A}, \qquad
    p_2=-(j^2-2)\fr{F_2r^2}{\sqrt{1-A}}, \qquad
    p_3=\fr{(F_2+q^2A_1)r^4}{\sqrt{1-A}}.
    \ee
We introduce an auxiliary variable $\chi$ to write~\cite{DeFelice:2011ka}
	\be
	\fr{2\ell+1}{2\pi j^2}\mL^{(2)}_{\rm odd}=
	p_1h_0^2+p_2h_1^2+p_3\brb{-\chi^2+2\chi \bra{\dot{h}_1-h'_0}}. \label{qlag2_odd}
	\ee
Note that the original quadratic Lagrangian~\eqref{qlag_odd} can be recovered by integrating out $\chi$.
Once written in the form~\eqref{qlag2_odd}, after integration by parts, $h_0$ and $h_1$ become auxiliary fields, so that their EOMs yield
	\be
	h_0=-\fr{(p_3\chi)'}{p_1}, \qquad
	h_1=\fr{(p_3\chi)^{\boldsymbol{\cdot}}}{p_2}. \label{h0h1}
	\ee
Then, the resubstitution into \eqref{qlag2_odd} yields the following Lagrangian written in terms of $\chi$ only:
	\be
	\fr{(j^2-2)(2\ell+1)}{2\pi j^2}\mL^{(2)}_{\rm odd}=
	s_1\dot{\chi}^2-s_2\chi'{}^2-\bra{j^2s_3+V}\chi^2, \label{qlag3_odd}
	\ee
where 
    \be
	s_1=\fr{(F_2+q^2A_1)^2r^6}{F_2\sqrt{1-A}}, \qquad
	s_2=\fr{(F_2+q^2A_1)r^6}{(1-A)^{3/2}}, \qquad
	s_3=\fr{(F_2+q^2A_1)r^4}{\sqrt{1-A}},
	\ee
and $V$ is written as
	\be
	V=s_3\brb{(s_2-s_1)\bra{\frac{1}{s_3}}'\,}'-2s_3.
	\ee
Note that one cannot rewrite the Lagrangian~\eqref{qlag2_odd} in the form~\eqref{qlag3_odd} for dipole perturbations with $\ell=1$ since $p_1=p_2=0$ in this particular case, and thus the denominators in \eqref{h0h1} vanish.
We shall address the case with $\ell=1$ in \S\ref{ssec:dipole_odd}.

From the Lagrangian~\eqref{qlag3_odd}, we see that there is no ghost/gradient instabilities if $s_1$, $s_2$, and $s_3$ are all positive, namely,
    \be
    F_2>0, \qquad
    F_2+q^2A_1>0. \label{stability_odd}
    \ee
This set of conditions is weaker than the one obtained in \cite{Takahashi:2019oxz}, where we obtained a third condition in addition to \eqref{stability_odd}. 
However, this is not a contradiction as the (un)boundedness of a Hamiltonian is coordinate-dependent in general~\cite{Babichev:2017lmw,Babichev:2018uiw}.
In other words, the above analysis in the Lema\^{i}tre coordinates clarifies that the third condition obtained in \cite{Takahashi:2019oxz} is actually not necessary to be imposed.
It should also be noted that the condition~\eqref{stability_odd} is consistent with the one for tensor perturbations about cosmological background~\cite{deRham:2016wji}.
Also, the squared radial sound speed~$c_\rho^2$ can be read off from \eqref{qlag3_odd} as
    \begin{equation}
    c_\rho^2=\fr{g_{\rho\rho}}{|g_{\tau\tau}|}\fr{s_2}{s_1}
    =\fr{F_2}{F_2+q^2A_1}\eqqcolon c_{\rm GW}^2. \label{ss_odd}
    \end{equation}
Here, the factor~$g_{\rho\rho}/|g_{\tau\tau}|$ is inserted so that a massless canonical scalar field propagating in the $\rho$-direction has $c_\rho^2=1$ in the natural units.
One can also compute the squared sound speed in the angular direction~$c_\theta^2$, noting that the factor~$j^2$ in front of the term with $s_3$ in \eqref{qlag3_odd} originates from the spherical Laplacian.
It is given by
    \be
    c_\theta^2=\fr{r^2}{|g_{\tau\tau}|}\fr{s_3}{s_1}
    =\fr{F_2}{F_2+q^2A_1}=c_{\rm GW}^2,
    \ee
where the factor~$r^2/|g_{\tau\tau}|$ plays the same role as $g_{\rho\rho}/|g_{\tau\tau}|$ in \eqref{ss_odd}.
This coincides with the squared sound speed in the radial direction.

\subsection{\texorpdfstring{Dipole perturbation:~$\ell=1$}{Dipole}}
\label{ssec:dipole_odd}

Let us now focus on the dipole perturbations with $\ell=1$,
for which $h_2$ is intrinsically absent. 
Up to \eqref{qlag_odd}, we can follow the same procedure for the case with $\ell\geq 2$. 
Plugging $\ell=1$, we have
	\be
	\fr{3}{2\pi}\mL^{(2)}_{\rm odd}=p_3\bra{\dot{h}_1-h'_0}^2. \label{qlag_dipole_odd}
	\ee
As we discussed above, for the dipole perturbations, there are no complete gauge fixing.
In this case, we should impose an incomplete gauge-fixing condition after deriving EL equations, rather than imposing 
it at the Lagrangian level~\cite{Motohashi:2016prk}. 
From the Lagrangian~\eqref{qlag_dipole_odd}, the EL equations are given by 
\be \label{ELeq_dipole_odd} \brb{p_3\bra{\dot{h}_1-h'_0} }^{\boldsymbol{\cdot}} = 0, \qquad \brb{p_3\bra{\dot{h}_1-h'_0} }' = 0. \ee
With the gauge transformation~\eqref{gaugetrnsf_odd}, we can set $h_1$ to zero by choosing the gauge function such that $\Xi'=h_1$.
Note that this is an incomplete gauge fixing in the sense that the gauge function $\Xi$ is not fully determined.
Indeed, there exists a residual gauge DOF~$\Xi=\Xi(\tau)$, i.e., a shift of $h_0$ by $\tau$-dependent function.
After the gauge fixing, the EL equations~\eqref{ELeq_dipole_odd} read $\bra{p_3h_0'}^{\boldsymbol{\cdot}}=\bra{p_3h_0'}'=0$, which implies that $p_3h_0'$ is constant both in $\tau$ and $\rho$.
Noting that $p_3^{-1}\propto (r^{-3})'$, the solution for $h_0$ can be written as
\be
h_0=-\fr{J}{8\pi (F_2+q^2A_1)r^3}, \label{odd_dipole_sol}
\ee
with $J$ being an arbitrary constant.
Note that a $\tau$-dependent integration constant has been set to zero by use of the residual gauge DOF mentioned above.
This result is consistent with the one in \cite{Takahashi:2019oxz}.
As was shown in \cite{Kobayashi:2012kh,Ogawa:2015pea}, the dipole modes of odd-parity perturbations are related to the slow rotation of a BH and the constant~$J$ amounts to its angular momentum, which is the reason for the contrived choice of the integration constant in \eqref{odd_dipole_sol}.

Before closing the odd-parity analysis, here is a good place to provide a specific example of 
incorrect gauge fixing at the Lagrangian level.
If one sets $h_1$ to zero at the Lagrangian~\eqref{qlag_dipole_odd}, one would obtain $\bra{p_3h_0'}'=0$ only and lose the other EL equation $\bra{p_3h_0'}^{\boldsymbol{\cdot}}=0$.
As a result, we do not arrive at \eqref{odd_dipole_sol} even after using the residual gauge DOF.
As shown in \cite{Motohashi:2016prk}, in general an incomplete gauge fixing at the Lagrangian level leads to a loss of independent EL equations and leads to an incorrect analysis.
Therefore, an incomplete gauge fixing should be imposed only after deriving all the EL equations, as we have done above.
On the other hand, a complete gauge fixing can be imposed at the Lagrangian level since the 
EL equations lost by the gauge fixing are redundant and can be recovered from the remaining EL equations via the Noether identity.

\section{Even-parity perturbations}\label{sec:even}

As we saw in the previous section, three out of ten components of the metric perturbation~$h_\mn=g_\mn-g^{(0)}_\mn$ belong to the odd modes.
In this section, we study the even modes, which are composed of the remaining seven components of the metric perturbation and the perturbation of the scalar field~$\delta\phi\coloneqq \phi-\phi^{(0)}$.
The even-parity perturbations of the metric are decomposed as follows:
	\be
	\begin{split}
	h_{\tau \tau }&=\sum_{\ell,m}H_{0,\ell m}(\tau ,\rho )Y_{\ell m}(\theta,\varphi),\\
	h_{\tau \rho }&=\sum_{\ell,m}H_{1,\ell m}(\tau ,\rho )Y_{\ell m}(\theta,\varphi),\\
	h_{\rho \rho }&=\brb{1-A(r)}\sum_{\ell,m}H_{2,\ell m}(\tau ,\rho )Y_{\ell m}(\theta,\varphi),\\
	h_{\tau a}&=\sum_{\ell,m}r^2\alpha_{\ell m}(\tau ,\rho )\bar{\na}_aY_{\ell m}(\theta,\varphi),\\
	h_{\rho a}&=\sum_{\ell,m}r^2\beta_{\ell m}(\tau ,\rho )\bar{\na}_aY_{\ell m}(\theta,\varphi),\\
	h_{ab}&=\sum_{\ell,m}r^2\brb{K_{\ell m}(\tau ,\rho )\ga_{ab}Y_{\ell m}(\theta,\varphi)+G_{\ell m}(\tau ,\rho )\bar{\na}_a\bar{\na}_bY_{\ell m}(\theta,\varphi)}.
	\end{split} \label{pert_even}
	\ee
Again, we factor out $r^2$ from the definition of $\alpha$, $\beta$, $K$, and $G$ for later convenience.
As for the perturbation of the scalar field, we write
	\be
	\delta\phi=\sum_{\ell,m}\delta\phi_{\ell m}(\tau,\rho)Y_{\ell m}(\theta,\varphi).
	\ee
The even-parity part of an infinitesimal coordinate transformation~$x^\mu\to x^\mu+\ep^\mu$ can be written as
	\be
	\ep^\tau=\sum_{\ell,m}T_{\ell m}(\tau,\rho)Y_{\ell m}(\theta,\varphi), \qquad
	\ep^\rho=\sum_{\ell,m}P_{\ell m}(\tau,\rho)Y_{\ell m}(\theta,\varphi), \qquad
	\ep^a=\sum_{\ell,m}\Theta_{\ell m}(\tau,\rho)\bar{\na}^aY_{\ell m}(\theta,\varphi).
	\ee
Correspondingly, the gauge transformation of the even-parity perturbations is given by
	\begin{align}
	H_0&\to H_0+2\dot{T}, &
	H_1&\to H_1+T'-(1-A)\dot{P}, &
	H_2&\to H_2+\fr{A'}{1-A}(P-T)-2P', \nonumber \\
	\alpha&\to \alpha+\fr{T}{r^2}-\dot{\Theta}, &
	\beta&\to \beta-\fr{1-A}{r^2}P-\Theta', &
	K&\to K-\fr{2r'}{r}(P-T), \label{gaugetrnsf_even} \\
	G&\to G-2\Theta, &
	\delta\phi&\to \delta\phi-qT. &
	\nonumber
	\end{align}

Let us consider appropriate gauge fixing for the even modes.
In the case of $\ell\ge 2$, the gauge-fixing condition~$K=G=\delta\phi=0$ determines all the gauge functions $T$, $P$, $\Theta$ without ambiguities.
Hence, it is a complete gauge fixing and can be imposed at the Lagrangian level~\cite{Motohashi:2016prk}.
We shall use this gauge fixing for $\ell\ge 2$ in \S\ref{ssec:multipole_even}.
However, it should be noted that the above discussion does not apply to monopole~($\ell=0$) and dipole~($\ell=1$) perturbations.
For the monopole perturbations, the variables~$\alpha$, $\beta$, and $G$, and also the gauge function~$\Theta$ are intrinsically absent.
Hence, we instead make a complete gauge fixing~$K=\delta\phi=0$ by choosing the remaining gauge functions~$T$ and $P$ appropriately, which we shall adopt in \S\ref{ssec:monopole}.
Regarding the dipole perturbations, due to the identity~$\bar{\na}_a\bar{\na}_bY_{\ell m}=-\gamma_{ab}Y_{\ell m}$ for $\ell=1$, we should treat the linear combination~$K-G$ as a single perturbation variable associated with $h_{ab}$.
We can then choose the gauge functions~$T$ and $\Theta$ so as to set $K-G=\delta\phi=0$, which can be imposed at the Lagrangian level.
Note that there exists a residual gauge DOF associated with the undetermined gauge function~$P$.
In \S\ref{ssec:dipole_even}, we specify a master variable which is invariant under this residual gauge transformation.

Before proceeding to the detailed analysis for each case, let us summarize our strategy for studying general multipole perturbations in \S\ref{ssec:multipole_even}.
Similarly to the case of odd modes, we first construct the quadratic Lagrangian for the even modes, which is written in terms of five perturbation variables left after the gauge fixing mentioned above.
We then perform a field redefinition to remove several terms with higher derivatives.
It should be noted that this field redefinition is related to a conformal/disformal transformation that maps the given DHOST theory into the Horndeski subclass, as we shall discuss in Appendix~\ref{AppA}.
Next, we make a further field redefinition, after which we can employ the same trick as used in \eqref{qlag2_odd} for the odd modes.
Namely, we complete the square of terms containing derivatives, and then introduce a new auxiliary variable and perform integration by parts to shift the derivatives to the new variable.
After this manipulation, we can integrate out remaining nondynamical variables simultaneously and find a quadratic Lagrangian~\eqref{qlag5_even} written in terms only of two master variables~$\si$ and $\zeta$, which we shall define in \eqref{sigma} and \eqref{zeta}, respectively. 
On the other hand, as mentioned earlier, the monopole and dipole perturbations should be treated separately.
We shall discuss these cases in \S\ref{ssec:monopole} and \S\ref{ssec:dipole_even}, respectively.
As a complementary analysis, we present a Hamiltonian analysis for the $\ell\ge 1$ case in Appendix~\ref{AppB} and a characteristic analysis for the $\ell=0$ case in Appendix~\ref{AppC}.

\subsection{\texorpdfstring{Even-parity perturbations with $\ell\ge 2$}{Multipole}}
\label{ssec:multipole_even}

Let us first consider higher multipoles with $\ell\ge 2$. 
As we did for the odd modes, we expand the perturbations in terms of the Legendre polynomials rather than the spherical harmonics.
After performing the integration over angular variables, the quadratic action for the even modes takes the form
	\be
	S^{(2)}_{\rm even}=\int \D\tau \D\rho \, \mL^{(2)}_{\rm even},
	\ee
where
	\begin{align}
	\fr{2\ell+1}{2\pi}\mL^{(2)}_{\rm even}=&~
	a_0\dot{H}'_0H_1+a_1\dot{H}_0^2+a_2\dot{H}_0\dot{H}_2+a_3H'_0{}^2+j^2a_4\bra{\dot{\beta}-\alpha'}^2+j^2b_1\dot{H}_0\alpha \nonumber \\
	&+b_2\dot{H}_2H_0+b_3\dot{H}_2H_1+j^2b_4\dot{H}_2\alpha
	+j^2b_5\dot{\beta}H_1
	+b_6H'_0H_1+b_7H'_0H_2+j^2b_8H'_0\beta+j^2b_9\alpha'H_1 \nonumber \\
	&+\bra{c_1+j^2c_2}H_0^2+j^2c_3H_0H_2+j^2c_4H_0\alpha+j^2c_5H_0\beta+j^2c_6H_1^2+j^2c_7H_1\alpha+j^2c_8H_1\beta \nonumber \\
	&+c_9H_2^2+j^2c_{10}H_2\alpha+j^2c_{11}H_2\beta+j^2c_{12}\alpha^2+j^2c_{13}\beta^2. \label{qlag_even}
	\end{align}
While we do not write down the explicit form of the coefficients here, note that the coefficients~$a_0$, $a_1$, $a_2$, $a_3$, $b_1$, and $c_2$ are proportional to the following quantity (or its square):
    \be
    \Up\coloneqq \fr{q^2(4F_{2X}-2A_1-q^2A_3)}{8(F_2+q^2A_1)}, \label{Upsilon}
    \ee
which vanishes for the case of Horndeski theories having $A_1=2F_{2X}$ and $A_3=0$.
The terms with these coefficients can be absorbed into the others by the following redefinition of $H_1$, $H_2$, and $\beta$:
    \be
    \begin{split}
    \ti{H}_1&\coloneqq H_1+(1-A)\bra{\fr{a_2}{(1-A)b_3}H_0}^{\boldsymbol{\cdot}} =H_1+\Up(1-A)\bra{\fr{r}{\sqrt{1-A}}H_0}^{\boldsymbol{\cdot}}, \\
    \ti{H}_2&\coloneqq H_2+\fr{a_0}{b_3}H'_0
    =H_2+2\Up\fr{r}{\sqrt{1-A}}H_0', \\
    \ti{\beta}&\coloneqq \beta+\fr{a_2}{r^2b_3}H_0
    =\beta+\Up\fr{\sqrt{1-A}}{r}H_0.
    \end{split} \label{map2Horndeski}
    \ee
This is as it should be since a class Ia DHOST theory can be mapped into the Horndeski class via redefinition of the metric, known as the conformal/disformal transformation (see Appendix~\ref{AppA}).
With the new variables, the quadratic Lagrangian is written as
	\begin{align}
	\fr{2\ell+1}{2\pi}\mL^{(2)}_{\rm even}=&~
	j^2\ti{a}_4\bra{\dot{\ti{\beta}}-\alpha'}^2
	+\ti{b}_2\dot{\ti{H}}_2H_0+\ti{b}_3\dot{\ti{H}}_2\ti{H}_1+j^2\ti{b}_4\dot{\ti{H}}_2\alpha
	+j^2\ti{b}_5\dot{\ti{\beta}}\ti{H}_1 \nonumber \\
	&+\ti{b}_6H'_0\ti{H}_1+\ti{b}_7H'_0\ti{H}_2+j^2\ti{b}_8H'_0\ti{\beta}+j^2\ti{b}_9\alpha'\ti{H}_1 \nonumber \\
	&+\ti{c}_1H_0^2+j^2\ti{c}_3H_0\ti{H}_2+j^2\ti{c}_4H_0\alpha+j^2\ti{c}_5H_0\ti{\beta}+j^2\ti{c}_6\ti{H}_1^2+j^2\ti{c}_7\ti{H}_1\alpha+j^2\ti{c}_8\ti{H}_1\ti{\beta} \nonumber \\
	&+\ti{c}_9\ti{H}_2^2+j^2\ti{c}_{10}\ti{H}_2\alpha+j^2\ti{c}_{11}\ti{H}_2\ti{\beta}+j^2\ti{c}_{12}\alpha^2+j^2\ti{c}_{13}\ti{\beta}^2. \label{qlag2_even}
	\end{align}
For convenience, we define the following functions: 
    \be
    \begin{split}
    \Psi(X)&\coloneqq 2F_0-2XF_{0X}+X^2F_{0XX}-X^2F_0\fr{(F_2-XA_1)_{XX}}{F_2-XA_1}+\fr{X^3}{F_2-XA_1}\bra{\fr{F_0\Phi}{X^2}}_X, \\
    \Phi(X)&\coloneqq F_2-XF_{2X}-\fr{3}{2}XA_1-2X^2A_{1X}-\fr{3}{4}X^2A_3, \\
    \Pi(X)&\coloneqq F_2A_3+2(F_2A_1)_X.
    \end{split}
    \ee
In terms of these functions, the coefficients in \eqref{qlag2_even} are expressed as
    \be
    \begin{split}
    &\ti{a}_4=\fr{(F_2+q^2A_1)r^4}{\sqrt{1-A}}, \quad
    \ti{b}_2=2\Phi r(1-A), \quad
    \ti{b}_3=4(F_2+q^2A_1)r, \quad
    \ti{b}_4=-2(F_2+q^2A_1)r^2\sqrt{1-A}, \quad \\
    &\ti{b}_5=-\fr{2(F_2+q^2A_1)r^2}{\sqrt{1-A}}, \quad
    \ti{b}_6=4\Phi r, \quad
    \ti{b}_7=\fr{2(F_2\Phi+q^4\Pi)r}{F_2+q^2A_1}, \quad
    \ti{b}_8=-\fr{2(F_2\Phi+q^4\Pi)r^2}{(F_2+q^2A_1)\sqrt{1-A}}, \quad 
    \ti{b}_9=\ti{b}_5, \\
    &\ti{c}_1=\bra{\Psi+\fr{F_0\Phi^2}{2(F_2+q^2A_1)^2}}r^2\sqrt{1-A}, \quad
    \ti{c}_3=-\fr{(F_2\Phi+q^4\Pi)\sqrt{1-A}}{F_2+q^2A_1}, \quad
    \ti{c}_4=2\Phi r\bra{r\sqrt{1-A}}', \\
    &\ti{c}_5=\ti{b}_7, \quad
    \ti{c}_6=\fr{F_2+q^2A_1}{\sqrt{1-A}}, \quad
    \ti{c}_7=-\ti{b}_3, \quad
    \ti{c}_8=4(F_2+q^2A_1)r\bra{\fr{r}{\sqrt{1-A}}}', \quad
    \ti{c}_9=F_2\sqrt{1-A}, \\
    &\ti{c}_{10}=2(F_2+q^2A_1)r^3\bra{\fr{\sqrt{1-A}}{r}}', \quad
    \ti{c}_{11}=-2F_2r, \quad
    \ti{c}_{12}=\ti{b}_4, \quad
    \ti{c}_{13}=\fr{2F_2r^2}{\sqrt{1-A}},
    \end{split} \label{coeffs}
    \ee
where $\Psi,\Phi,\Pi$ are evaluated at $X=-q^2$, same as the coefficient functions.

In general, the Lagrangian~\eqref{qlag2_even} yields two dynamical DOFs, one of which corresponds to gravitational waves and the other corresponds to scalar waves.
However, when $\Psi=0$ (at least at $X=-q^2$), a Hamiltonian analysis shows that there is only one dynamical DOF (see Appendix~\ref{AppB}). 
Such a situation occurs either because the system has two DOFs but one of them is strongly coupled, or the system has inherently one DOF. 
The latter case applies to theories where scalar waves do not propagate~\cite{Lin:2017oow,Chagoya:2018yna,Aoki:2018zcv,Afshordi:2006ad,Iyonaga:2018vnu,Iyonaga:2020bmm,Gao:2019twq}.
As a general class of theories with this property, we consider the extended cuscutons constructed in \cite{Iyonaga:2018vnu}.
The shift- and reflection-symmetric subclass of extended cuscutons up to quadratic-order interactions is embedded in \eqref{qDHOST} as
    \be
    F_0=u_0+u_1\sqrt{-X},\qquad
    F_2=u_2+u_3\sqrt{-X},\qquad
    A_3=\fr{1}{X^2}\bra{2u_2+\fr{u_4}{1+u_5\sqrt{-X}}},\qquad
    A_1=2F_{2X}+\fr{X}{2}A_3, \label{cuscuton}
    \ee
with $u_i$'s being constant.
In order for this class of theories to satisfy the condition~\eqref{exist}, one has to tune the coefficient~$u_1$ as $u_1=u_0u_5$.
Also, $u_4\ne 0$ is necessary for $F_2-XA_1\ne 0$.
Then, it is straightforward to check $\Psi(X)=0$.
As such, $\Psi=0$ results in the absence of scalar waves.
For the following, we restrict ourselves to $\Psi\ne 0$. 
Moreover, as we shall see later, either $\Phi=0$ or $\Pi=0$ leads to a vanishing sound speed for scalar waves, which signals the strong coupling.
Therefore, in what follows, we assume $\Phi\ne 0$ and $\Pi\ne 0$ unless otherwise stated (see Appendices~\ref{AppB} and \ref{AppC} for the exceptional cases where any of $\Psi$, $\Phi$, or $\Pi$ is vanishing).

To further reduce the Lagrangian, we introduce a new variable~$\zeta$ by
    \be
    \zeta\coloneqq 
    \ti{H}_1+\fr{\ti{b}_2}{\ti{b}_3}H_0+\fr{\ti{b}_7}{\ti{b}_6}\ti{H}_2
    +j^2\fr{\ti{b}_4}{\ti{b}_3}\alpha+j^2\fr{\ti{b}_8}{\ti{b}_6}\ti{\beta}. \label{zeta}
    \ee
In terms of $\zeta$ instead of $\ti{H}_1$, the quadratic Lagrangian~\eqref{qlag2_even} is rewritten in the following form:
    \begin{align}
	\fr{2\ell+1}{2\pi}\mL^{(2)}_{\rm even}=&~
	j^2\ti{a}_4\brb{\dot{\ti{\beta}}-\alpha'
	+\fr{1}{r^2}\bra{\fr{\ti{b}_2}{\ti{b}_3}H_0-\fr{\ti{b}_7}{\ti{b}_6}\ti{H}_2
    +j^2\fr{\ti{b}_4}{\ti{b}_3}\alpha-j^2\fr{\ti{b}_8}{\ti{b}_6}\ti{\beta}}}^2 \nonumber \\
    &+\zeta\bra{\ti{b}_3\dot{\ti{H}}_2+\ti{b}_6H'_0+j^2\ti{b}_5\dot{\ti{\beta}}+j^2\ti{b}_9\alpha'}
    +j^2\ti{c}_6\zeta^2+\sum_{i=1}^4\mV_iQ^i\zeta+\fr{1}{2}\sum_{i,j=1}^4\mU_{ij}Q^iQ^j,
    \label{qlag3_even}
    \end{align}
with $Q^i\coloneqq(H_0,\ti{H}_2,\alpha,\ti{\beta})$.
Then, as we did for the case of odd modes, we introduce an auxiliary variable~$\psi$ to rewrite the Lagrangian~\eqref{qlag3_even} in an equivalent form,
    \begin{align}
	\fr{2\ell+1}{2\pi}\mL^{(2)}_{\rm even}=&~
	j^2\ti{a}_4\brc{-\psi^2+2\psi\brb{\dot{\ti{\beta}}-\alpha'
	+\fr{1}{r^2}\bra{\fr{\ti{b}_2}{\ti{b}_3}H_0-\fr{\ti{b}_7}{\ti{b}_6}\ti{H}_2
    +j^2\fr{\ti{b}_4}{\ti{b}_3}\alpha-j^2\fr{\ti{b}_8}{\ti{b}_6}\ti{\beta}}}} \nonumber \\
    &+\zeta\bra{\ti{b}_3\dot{\ti{H}}_2+\ti{b}_6H'_0+j^2\ti{b}_5\dot{\ti{\beta}}+j^2\ti{b}_9\alpha'}
    +j^2\ti{c}_6\zeta^2+\sum_{i=1}^4\mV_iQ^i\zeta+\fr{1}{2}\sum_{i,j=1}^4\mU_{ij}Q^iQ^j. \label{qlag4_even}
    \end{align}
Indeed, the EL equation for $\psi$ yields
    \be
    \psi=
    \dot{\ti{\beta}}-\alpha'
	+\fr{1}{r^2}\bra{\fr{\ti{b}_2}{\ti{b}_3}H_0-\fr{\ti{b}_7}{\ti{b}_6}\ti{H}_2
    +j^2\fr{\ti{b}_4}{\ti{b}_3}\alpha-j^2\fr{\ti{b}_8}{\ti{b}_6}\ti{\beta}}, \label{psi}
    \ee
which can be resubstituted into \eqref{qlag4_even} to recover the Lagrangian~\eqref{qlag3_even}.
From \eqref{qlag4_even}, we see that $Q^i$'s are nondynamical variables and their EL equations take the following form:
\be \sum_{j=1}^4\mU_{ij}Q^j = (\text{terms with $\psi$, $\zeta$, and their first derivatives}) . \label{EOM_Qi} \ee
The determinant of the matrix~$\mU_{ij}$ is given by
    \be
    \det \mU=\bra{j^2\fr{F_2+q^2A_1}{\Phi }r^2}^2(j^2\mA^2+\mB_1)(j^2\mA^2+\mB_2), \label{detU}
    \ee
where
    \be
    \mA\coloneqq 2r^2\bra{\fr{\sqrt{1-A}}{r}}'\Phi -(j^2-2)\fr{F_2\Phi+q^4\Pi}{F_2+q^2A_1}, \quad
    \mB_1\coloneqq 4(j^2-2)\fr{q^4\Phi \Pi}{F_2+q^2A_1}(1-A), \quad
    \mB_2\coloneqq 4(j^2-2)F_2\Psi r^2. \label{AandBn}
    \ee
From the stability condition~\eqref{stability_odd} for odd modes, we have $F_2>0$ and $F_2+q^2A_1>0$.
Moreover, as we shall see in \S\ref{ssec:monopole}, the stability of monopole perturbations require $\Phi \Pi>0$ and $\Psi >0$.
These two facts ensure that $\det \mU$ is nonvanishing, and hence \eqref{EOM_Qi} can be solved algebraically to express $Q^i$'s in terms of $\psi$, $\zeta$, and their first derivatives.
Thus, we obtain a Lagrangian as a functional of $\psi$ and $\zeta$.
Finally, replacing $\psi$ by $\si\coloneqq r^2\psi+\zeta$, i.e.,
    \be
    \si=
    \ti{H}_1+\fr{2\ti{b}_2}{\ti{b}_3}H_0
    +2j^2\fr{\ti{b}_4}{\ti{b}_3}\alpha
    +r^2\bra{\dot{\ti{\beta}}-\alpha'}, \label{sigma}
    \ee
we arrive at the following expression: 
    \begin{align}
	\fr{2\ell+1}{2\pi}\mL^{(2)}_{\rm even}=&~
	\sum_{I,J=1}^2\bra{\fr{1}{2}\mK_{IJ}\dot{v}^I\dot{v}^J
	+\mM_{IJ}\dot{v}^Iv^{J\prime}
	-\fr{1}{2}\mG_{IJ}v^{I\prime}v^{J\prime}
	-\fr{1}{2}\mW_{IJ}v^Iv^J}.
	\label{qlag5_even}
	\end{align}
This is the quadratic Lagrangian for the even-parity perturbations written in terms of the two master variables $v^I\coloneqq (\si,\zeta)$, which are related to the metric perturbations via \eqref{sigma} and \eqref{zeta}, respectively. 
The nonvanishing components of the matrices~$\mK_{IJ}$, $\mM_{IJ}$, and $\mG_{IJ}$ are given by
    \be
    \begin{split}
    \mK_{11}&=8j^2(F_2+q^2A_1)^2\Psi \fr{r^4}{\sqrt{1-A}(j^2\mA^2+\mB_2)}, \\
    \mK_{22}&=32(j^2-2)(F_2+q^2A_1)\Phi ^2\fr{r^2\sqrt{1-A}}{j^2\mA^2+\mB_1}, \\
    \mM_{12}=\mM_{21}&=4j^2(F_2+q^2A_1)\Phi \fr{r^2\mA(\mB_2-\mB_1)}{\sqrt{1-A}(j^2\mA^2+\mB_1)(j^2\mA^2+\mB_2)}, \\
    \mG_{11}&=8j^2q^4\Phi \Pi\fr{r^2}{\sqrt{1-A}(j^2\mA^2+\mB_1)}, \\
    \mG_{22}&=32(j^2-2)F_2\Phi ^2\fr{r^2}{\sqrt{1-A}(j^2\mA^2+\mB_2)}.
    \end{split} \label{KMG}
    \ee
Since the explicit form of the matrix~$\mW_{IJ}$ is too lengthy, we just present the expression in the large-$\ell$ limit:
    \be
    \begin{split}
    \mW_{11}&=-4j^{-2}q^4\Phi\Pi\bra{\fr{F_2+q^2A_1}{F_2\Phi+q^4\Pi}}^2\fr{rA'}{1-A}+\mO(j^{-4}), \\
    \mW_{12}=\mW_{21}&=-4j^{-2}q^4\Phi\Pi\bra{\fr{F_2+q^2A_1}{F_2\Phi+q^4\Pi}}^2\fr{r^3}{1-A}\bra{\fr{1-A}{r^2}}'+\mO(j^{-4}), \\
    \mW_{22}&=32j^{-2}F_2\Phi^2\bra{\fr{F_2+q^2A_1}{F_2\Phi+q^4\Pi}}^2\sqrt{1-A}+\mO(j^{-4}),
    \end{split} \label{WIJ}
    \ee
which will be used to compute the squared sound speed in the angular direction.

Given that the stability condition~\eqref{stability_odd} for odd modes is satisfied, $\Phi \Pi>0$ and $\Psi >0$ guarantee $\mK_{11}>0$ and $\mK_{22}>0$, and hence the system is free of ghost instabilities.
Note that this condition coincides with the stability criterion~\eqref{stability_even} for monopole perturbations.
From \eqref{qlag5_even}, one can also obtain the squared radial sound speed~$c_\rho^2$ for the two DOFs as solutions for the following algebraic equation:
    \be
    \det \brb{c_\rho^2\mK_{IJ}+2\sqrt{\fr{g_{\rho\rho}}{|g_{\tau\tau}|}}\,c_\rho\mM_{IJ}-\fr{g_{\rho\rho}}{|g_{\tau\tau}|}\mG_{IJ}}=0. \label{eq_ss}
    \ee
This equation yields two solutions for $c_\rho^2$.
The first solution is given by
    \be
    c_\rho^2=\fr{F_2}{F_2+q^2A_1}=c_{\rm GW}^2,
    \ee
which coincides with $c_\rho^2$ for odd modes obtained in \eqref{ss_odd}.
Therefore, this first mode can be interpreted as gravitational waves.
On the other hand, the second solution for \eqref{eq_ss} is given by
    \be
    c_\rho^2=\fr{q^4\Phi \Pi}{\Psi (F_2+q^2A_1)^2}\fr{1-A}{r^2}\eqqcolon c_{\rm SW}^2, \label{ss_scalar}
    \ee
which can be identified as $c_\rho^2$ for scalar waves.
Indeed, as we shall see in the subsequent sections, the above $c_{\rm SW}^2$ coincides with the squared radial sound speed for the monopole and dipole perturbations.
Note that, although the coefficients in \eqref{eq_ss} depend on $\ell$, the above expressions for the squared radial sound speeds are independent of $\ell$.

It should be noted that the above discussion does not apply if any of $\Psi$, $\Phi$, and $\Pi$ vanishes, though the expression for $c_{\rm SW}^2$ seems to work at least for $\Phi=0$ and/or $\Pi=0$, where one finds $c_{\rm SW}^2=0$.
Remarkably, $\Pi=0$ happens even in the simplest case of the k-essence theory (i.e., GR plus $F_0(X)$~\cite{ArmendarizPicon:1999rj}).
Nevertheless, for such a case, it is reasonable to define $c_{\rm SW}^2=0$ as we shall see in Appendix~\ref{AppC} for the monopole case.

Next, let us study the squared sound speed in the angular direction~$c_\theta^2$. 
As mentioned in \S\ref{ssec:multipole_odd}, the factor~$j^2$ multiplied to the perturbation variables can be regarded as the spherical Laplacian, whose coefficients define $c_\theta^2$.
In the present case, the terms of $\mO(j^{-2})$ in \eqref{WIJ} correspond to the spherical Laplacian.
Indeed, in the large-$\ell$ limit, the coefficients of the kinetic term can be evaluated as
    \be
    \begin{split}
    \mK_{11}&=8j^{-4}\Psi\fr{(F_2+q^2A_1)^4}{(F_2\Phi+q^4\Pi)^2}\fr{r^4}{\sqrt{1-A}}+\mO(j^{-6}), \\
    \mK_{22}&=32j^{-4}\Phi^2\fr{(F_2+q^2A_1)^3}{(F_2\Phi+q^4\Pi)^2}r^2\sqrt{1-A}+\mO(j^{-6}).
    \end{split}
    \ee
Namely, if we rescale the fields as $\ti{v}^I=j^{-2}v^I$, we have $\mK_{IJ}=\mO(j^0)$ and $\mW_{IJ}=\mO(j^2)$, and hence it is reasonable to identify the leading coefficients of $\mW_{IJ}$ correspond to the spherical Laplacian.
From this observation, the squared angular sound speed~$c_\theta^2$ can be obtained as solutions for the following equation:
    \be
    \det \brb{\lim_{\ell\to\infty}\bra{c_\theta^2j^4\mK_{IJ}-\fr{r^2}{|g_{\tau\tau}|}j^{2}\mW_{IJ}}}
    =\frac{ 256 \Phi^2 \Psi (F_2+q^2 A_1)^7 r^6 }{ (F_2 \Phi+q^4 \Pi)^4 }\mF(c_\theta^2)
    =0, \label{ss_angular_eq}
    \ee
where the function~$\mF$ is a quadratic polynomial given by
    \be
    \mF(x)=\bra{x-c_{\rm GW}^2}\bra{x+\fr{rA'}{2(1-A)}c_{\rm SW}^2}
    -\fr{\Psi(F_2+q^2A_1)}{16\Phi^2}\bra{\fr{r^2(c_{\rm SW}^2)'}{1-A}}^2.
    \ee
Note that the last term is negative since $\Psi(F_2+q^2A_1)>0$, which is guaranteed from the stability conditions for odd modes [see \eqref{stability_odd}] and monopole perturbations [see \eqref{stability_even} in the next section].
Also, typically we have $c_{\rm GW}^2>-[rA'/2(1-A)]c_{\rm SW}^2$.
Hence, the equation~$\mF(x)=0$ has two real solutions, where one of the solutions is larger than $c_{\rm GW}^2$ and the other one is less than $-[rA'/2(1-A)]c_{\rm SW}^2$.
Let us call these solutions~$c_{\theta,+}^2$ and $c_{\theta,-}^2$, respectively.
Then, $c_{\theta,+}^2$, which is larger than $c_{\rm GW}^2$, is trivially positive.
On the other hand, in the vicinity of the BH horizon where $A'>0$, we have $c_{\theta,-}^2<-[rA'/2(1-A)]c_{\rm SW}^2<0$.
The existence of a negative solution implies that even-parity perturbations with $\ell\ge 2$ exhibit gradient instabilities in the angular direction, as was conjectured in \cite{Khoury:2020aya} for the stealth Schwarzschild black holes in Horndeski theories.
Conversely, if we try to make $c_{\theta,-}^2>0$, we need to flip the sign of $c_{\rm SW}^2$.
Hence, we have gradient instabilities either in the radial or angular direction at least in the vicinity of the BH horizon.
Nevertheless, so long as the value of $c_{\theta,-}^2$ is tiny, we expect it can be made positive by the scordatura effect~\cite{Motohashi:2019ymr}.
We shall discuss this point in detail in \S\ref{ssec:strong_coupling}.

\subsection{\texorpdfstring{Monopole perturbation:~$\ell=0$}{Monopole}}
\label{ssec:monopole}

Since the determinant~\eqref{detU} vanishes for $\ell=0$, we need to study the monopole perturbation separately.
Let us go back to the quadratic Lagrangian~\eqref{qlag3_even}.
By taking $\ell=0$ in \eqref{qlag3_even}, we obtain the following quadratic Lagrangian for monopole perturbations:
    \be
    \fr{1}{2\pi}\mL^{(2)}_{\rm even}=
    \zeta\bra{d_1\dot{\ti{H}}_2+d_2H'_0}+\fr{d_3}{2}H_0^2+\fr{d_4}{2}\ti{H}_2^2, \label{qlag3_mono}
    \ee
where
    \be
    \begin{split}
    &d_1\coloneqq \ti{b}_3=4r(F_2+q^2A_1), \qquad
    d_2\coloneqq \ti{b}_6=4\Phi r, \\
    &d_3\coloneqq 2\ti{c}_1+\bra{\fr{\ti{b}_2\ti{b}_6}{\ti{b}_3}}'=2\Psi r^2\sqrt{1-A}, \qquad
    d_4\coloneqq 2\ti{c}_9+\bra{\fr{\ti{b}_3\ti{b}_7}{\ti{b}_6}}^{\boldsymbol{\cdot}}=-\fr{2q^4\Pi \sqrt{1-A}}{\Phi}.
    \end{split}
    \ee
From \eqref{qlag3_mono}, one can integrate out $H_0$ and $\ti{H}_2$ to obtain the following quadratic Lagrangian for $\zeta$:
    \be
    \fr{1}{2\pi}\mL^{(2)}_{\rm even}=\fr{e_1}{2}\dot{\zeta}^2-\fr{e_2}{2}\zeta'^2-\fr{e_3}{2}\zeta^2. \label{qlag_monopole}
    \ee
The coefficients are written as
    \be
    e_1=\fr{8\Phi (F_2+q^2A_1)^2}{q^4\Pi}\fr{r^2}{\sqrt{1-A}}, \qquad
    e_2=\fr{8\Phi ^2}{\Psi \sqrt{1-A}}, \qquad
    e_3=\fr{2(1-A)}{r^2}e_2.
    \ee

The condition for the absence of ghost/gradient instabilities is given by $e_1>0$ and $e_2>0$, namely,
    \be
    \Phi \Pi>0, \qquad
    \Psi >0. \label{stability_even}
    \ee
Also, from the quadratic Lagrangian~\eqref{qlag_monopole}, the squared radial sound speed~$c_\rho^2$ for monopole perturbations is given as
    \be
    c_\rho^2=\fr{g_{\rho\rho}}{|g_{\tau\tau}|}\fr{e_2}{e_1}
    =\fr{q^4\Phi \Pi}{\Psi (F_2+q^2A_1)^2}\fr{1-A}{r^2}=c_{\rm SW}^2, \label{ss_mono}
    \ee
which coincides with \eqref{ss_scalar}.

\subsection{\texorpdfstring{Dipole perturbation:~$\ell=1$}{Dipole}}
\label{ssec:dipole_even}

As explained earlier, for dipole perturbations, one can fix $K-G=\delta\phi=0$ by choosing the gauge functions~$T$ and $\Theta$ appropriately, and hence obtains the same quadratic Lagrangian~\eqref{qlag5_even} as for multipole perturbations with $\ell\ge 2$.
When one performs a gauge transformation associated with the remaining gauge function~$P$, a transformation associated with $\Theta$ should also be performed so as to maintain the gauge condition~$K-G=0$.
More concretely, one must perform a gauge transformation associated with both $P$ and $\Theta$, where $\Theta$ is related to $P$ via $\Theta=(r'/r)P$. 
One can show that $\si$, one of the master variables for general multipole perturbations, is invariant under this residual gauge transformation.
Hence, as it should be, setting $\ell=1$ in \eqref{qlag5_even}, all the terms containing the other master variable~$\zeta$ vanish and we are left with the following Lagrangian written in terms of $\si$ only:
    \be
    \fr{3}{2\pi}\mL^{(2)}_{\rm even}=\fr{f_1}{2}\dot{\si}^2-\fr{f_2}{2}\si'^2-\fr{f_3}{2}\si^2, \label{qlag_dipole}
    \ee
where
    \be
    \begin{split}
    &f_1=\mK_{11}=\fr{2\Psi (F_2+q^2A_1)^2}{\Phi ^2\sqrt{1-A}}\brb{\bra{\fr{\sqrt{1-A}}{r}}'\,}^{-2}, \qquad
    f_2=\mG_{11}=\fr{2q^4\Pi}{\Phi  r^2\sqrt{1-A}}\brb{\bra{\fr{\sqrt{1-A}}{r}}'\,}^{-2}, \\
    &f_3=\fr{4(1-A)}{r^2}f_1+\fr{r^2}{1-A}\brb{\bra{\fr{\sqrt{1-A}}{r}}'\,}^2(f_1-f_2).
    \end{split} \label{qlag_dipole_coeff}
    \ee
In other words, one can regard that the residual gauge DOF is used to fix $\zeta$.  
Indeed, since $\zeta$ is transformed as
    \be
    \zeta\to\zeta-\fr{1}{2r^2}\bra{\fr{1-A}{r^2}}'P,
    \ee
one can make a complete gauge fixing by imposing, say, $\zeta=0$.
Here, the choice of the value of $\zeta$ is not important, since, as mentioned above, the quadratic Lagrangian for the dipole perturbations does not contain $\zeta$.
Therefore, in any case, we end up with the same quadratic Lagrangian~\eqref{qlag_dipole} after the complete gauge fixing.

From \eqref{qlag_dipole_coeff}, we see that there is no ghost/gradient instabilities so long as the condition~\eqref{stability_even} is satisfied.
Also, the squared radial sound speed for dipole perturbations is
    \be
    c_\rho^2=\fr{g_{\rho\rho}}{|g_{\tau\tau}|}\fr{f_2}{f_1}
    =\fr{q^4\Phi \Pi}{\Psi (F_2+q^2A_1)^2}\fr{1-A}{r^2}=c_{\rm SW}^2.
    \ee
As was the case of monopole perturbations, this expression coincides with \eqref{ss_scalar}.

\subsection{Strong coupling and gradient instability}\label{ssec:strong_coupling}

With the radial and angular sound speeds obtained above, let us study the strong coupling and gradient instability.
It was shown in \cite{Motohashi:2019ymr} that perturbations about stealth solutions in general DHOST theories have a vanishing sound speed squared in the asymptotic de~Sitter (or Minkowski) region.\footnote{In the presence of the $F_1(X)\Box\phi$ term in the action, gradient instability shows up, but in a rather narrow window at low energies~\cite{Motohashi:2019ymr}.
Since we set $F_1=0$ in our action~\eqref{qDHOST}, this gradient instability does not appear in the present case.}
The vanishing sound speed implies that the perturbations would be strongly coupled and hence the linear perturbation analysis can no longer be trusted.
This observation comes from the fact that the terms cubic in perturbations typically contain an inverse power of the sound speed in their coefficients in the language of the EFT of inflation~\cite{Cheung:2007st}, and therefore the strong coupling scale would be much lower than $M$ when the sound speed is vanishing.

As we see below, our result is consistent with the one in \cite{Motohashi:2019ymr}.
Substituting the explicit form of the function~$A(r)$, the squared radial sound speed for scalar waves can be expressed as
    \be
    c_{\rm SW}^2=\fr{q^4\Phi \Pi}{\Psi (F_2+q^2A_1)^2}\brb{\fr{\mu}{r^3}+\fr{|F_0|}{6(F_2+q^2A_1)}}.
    \ee
Here, we recall that the effective cosmological constant is given by \eqref{Lambda}, and we require $F_0\le 0$ so that $\Lambda\ge 0$.
Let us introduce a mass scale~$M$ and dimensionless quantities as $X(=-q^2)\eqqcolon M^4\hat{X}$, $F_0\eqqcolon M^4\hat{F}_0$, $A_1\eqqcolon M^{-2}\hat{A}_1$, and $A_3\eqqcolon M^{-6}\hat{A}_3$, whereas the function~$F_2$ [i.e., the coefficient in front of the Ricci scalar in \eqref{qDHOST}] is normalized as $F_2\eqqcolon \Mpl^2/2+M^2\hat{F}_2$, with $\Mpl$ being the reduced Planck mass.
The parameter~$M$, which we assume to be well below $\Mpl$, can be regarded as an energy scale at which higher-derivative terms of the scalar field come into play.
Then, provided that the ``hatted'' dimensionless quantities are of order unity, we have 
    \be
    c_{\rm SW}^2=x_0\bra{\fr{r_{0}}{r}}^3+x_1\bra{\fr{M}{\Mpl}}^2, \label{ss_scalar_approx}
    \ee
where $x_0$ and $x_1$ are (at most) of order unity and $r_{0}\coloneqq (\mu/M^2)^{1/3}$.
The stability conditions~\eqref{stability_odd} and \eqref{stability_even} guarantee that $x_0$ and $x_1$ are positive. 
First, we find $c_{\rm SW}^2\ll 1$ far outside the radius~$r_{0}$, which implies the strong coupling and is consistent with the result of \cite{Motohashi:2019ymr}.
On the other hand, near the BH horizon~$r\sim \mu$, we have $c_{\rm SW}^2\sim (r_{0}/\mu)^3\lesssim 1$ as we assume $\mu^{-1}\lesssim M$ to guarantee the validity of the EFT.
In particular, for $r_{0}\ll \mu$, one has $c_{\rm SW}^2\ll 1$ near the BH horizon.
Hence, in this case, $c_{\rm SW}^2$ is tiny everywhere from the BH horizon to the asymptotic de Sitter (or Minkowski) region.
It should also be mentioned that there are some exceptional cases where the sound speed exactly vanishes (see the \hyperref[table]{Table} in Appendix~\ref{AppC3}), and hence the perturbations would be infinitely strongly coupled.
Related to this point, the authors of \cite{deRham:2019gha} derived a set of conditions for perturbations about solutions with constant $X$ in DHOST theories to be infinitely strongly coupled, which is also consistent with our result.

Regarding the squared angular sound speed~$c_\theta^2$ for even-parity perturbations, we can make a similar order estimate as above.
As explained in \S\ref{ssec:multipole_even}, we have two solutions~$c_\theta^2=c_{\theta,\pm}^2$ corresponding to the two DOFs, which can be respectively estimated as
    \be
    c_{\theta,+}^2=c_{\rm GW}^2+x_2\bra{\fr{M}{\Mpl}}^2, \qquad
    c_{\theta,-}^2=-\fr{x_0}{2}\bra{\fr{r_{0}}{r}}^3+x_3\bra{\fr{M}{\Mpl}}^2, \label{ss_angular_estimate}
    \ee
where $x_2,x_3$ are of order unity and $x_0(>0)$ is the same one that appeared in \eqref{ss_scalar_approx}.
Since $c_{\rm GW}^2=\mO(1)$, the first equation in \eqref{ss_angular_estimate} means that $c_{\theta,+}^2$ is of the same order as $c_{\rm GW}^2$.
The second equation for $c_{\theta,-}^2$ is rather nontrivial.
First, as was the case for $c_{\rm SW}^2$, we note that $|c_{\theta,-}^2|\ll 1$ far outside the radius~$r_{0}$, indicating the strong coupling.
On the other hand, near the BH horizon $r\sim \mu$, we have $c_{\theta,-}^2\sim -c_{\rm SW}^2/2$.
Therefore, either $c_{\rm SW}^2$ or $c_{\theta,-}^2$ is negative.
This implies that the gradient instability appears either in the radial or angular direction, as mentioned earlier in \S\ref{ssec:multipole_even}.

To summarize, we showed that the perturbations would be strongly coupled in the asymptotic de Sitter (or Minkowski) region. 
Also, we clarified that the gradient instability inevitably shows up in the vicinity of the BH.
It should be noted that one can choose the theory parameters to make both $c_{\rm SW}^2$ and $c_{\theta,-}^2$ tiny or even zero everywhere.
In such a case, we expect that a controlled detuning of the degeneracy (i.e., the scordatura effect~\cite{Motohashi:2019ymr}) can alleviate the problem of gradient instability/strong coupling.

\section{Conclusions}\label{sec:conc}

We studied perturbations about the stealth Schwarzschild-de~Sitter solution in the quadratic DHOST theories.
For both odd- and even-parity perturbations, we clarified the master variable(s) and derived the quadratic Lagrangian for them.
For the odd-parity perturbations, the set of stability conditions we obtained in the Lema{\^i}tre coordinates is weaker than the one in \cite{Takahashi:2019oxz}.
This is because in general the (un)boundedness of a Hamiltonian is coordinate-dependent.
Our analysis clarified that one of the known stability conditions is actually not necessary to be imposed.
For the even-parity perturbations, we elucidated two master variables for the first time. 
One of the two modes is identified as gravitational waves as it has the same sound speed as odd modes, while the other is identified as scalar waves.
Based on the quadratic action in terms of the master variables, we verified that the perturbations would be strongly coupled in the asymptotic region where the spacetime is effectively described by the de Sitter (or Minkowski) metric.
Moreover, we studied the sound speed in the vicinity of the BH to find that the gradient instability inevitably shows up either in the radial or angular direction.

Let us clarify the limitation of our analysis, and discuss several possible ways out and extensions.
It should be noted that our analysis is based on the Lema{\^i}tre coordinates and the situation could change in other coordinate systems~\cite{Babichev:2018uiw}.
Another caveat is that one can choose the theory parameters to make the sound speed for scalar waves tiny or even zero everywhere.
In such a case, a slight detuning of the degeneracy conditions can cure the problem of gradient instability/strong coupling, which is known as the scordatura mechanism~\cite{Motohashi:2019ymr}.
It is known that the scordatura is important on cosmological scales~\cite{Gorji:2020bfl}.
Since we observe BHs at cosmological distances, it would be intriguing to see how the scordatura effect shows up in observables, e.g., quasi-normal modes~\cite{Nakashi_prep}.
It is also of interest to discuss BH perturbations in theories without propagating scalar DOF~\cite{Lin:2017oow,Chagoya:2018yna,Aoki:2018zcv,Afshordi:2006ad,Iyonaga:2018vnu,Iyonaga:2020bmm,Gao:2019twq}, where the problem of gradient instability/strong coupling would be intrinsically absent.
Another thing to investigate is the stability of non-stealth and/or $X\ne {\rm const}$ solutions. 
The linear stability analysis for odd-parity perturbations about a general static and spherically symmetric vacuum spacetime with a general scalar field profile in DHOST theories was completed in \cite{Takahashi:2019oxz}, and a similar analysis for even-parity perturbations would be possible as an extension of the analysis done in the present paper.
We leave these issues for future work.


\acknowledgments{We thank Masashi Kimura, Shinji Mukohyama, and Keisuke Nakashi for enlightening discussions.
K.T.\ was supported by JSPS (Japan Society for the Promotion of Science) KAKENHI Grant No.\ JP21J00695.
H.M.\ was supported by JSPS KAKENHI Grant No.\ JP18K13565.
}


\appendix

\section{Perturbation variables in the conformal/disformal frame}\label{AppA}

In this appendix, we discuss how the perturbation variables in the Lema{\^i}tre coordinates are transformed under the conformal/disformal transformation~\cite{Bekenstein:1992pj},
    \be
    g_\mn\quad\to\quad\bar{g}_\mn=\Omega(X)g_\mn+\Gamma(X)\phi_\mu\phi_\nu. \label{disformal}
    \ee
This transformation is invertible so long as $\Omega(\Omega-X\Omega_X-X^2\Gamma_X)\ne 0$~\cite{Zumalacarregui:2013pma}, and hence does not change the number of physical DOFs~\cite{Domenech:2015tca,Takahashi:2017zgr,Babichev:2019twf}.
In fact, the class of DHOST theories is closed under the conformal/disformal transformation~\cite{BenAchour:2016fzp}.
As was shown in \cite{Takahashi:2019oxz}, the stealth Schwarzschild-de~Sitter solution~\eqref{stealthSdS} in the original frame is mapped to a solution of the same form.
Indeed, for the background metric~$g^{(0)}_\mn$ in \eqref{Lemaitre}, performing the transformation~\eqref{disformal} yields
    \begin{align}
    \bar{g}^{(0)}_{\mn}\D x^\mu \D x^\nu&=
    -(\Omega-q^2\Gamma)\D\tau^2+\Omega\brb{1-A(r)}\D\rho^2+\Omega r^2\ga_{ab}\D x^a\D x^b \nonumber \\
    &=-\D\bar{\tau}^2+\brb{1-\bar{A}(\bar{r})}\D\bar{\rho}^2+\bar{r}^2\ga_{ab}\D x^a\D x^b, \label{disformal_SdS}
    \end{align}
where $\bar{\tau}=\sqrt{\Omega-q^2\Gamma}\,\tau$, $\bar{\rho}=\sqrt{\Omega-q^2\Gamma}\,\rho$, $\bar{r}=\sqrt{\Omega}\,r$, and
    \begin{align}
    \bar{A}(\bar{r})=\fr{\Omega A(\bar{r}/\sqrt{\Omega})-q^2\Gamma}{\Omega-q^2\Gamma}
    =1-\fr{\Omega^{3/2}}{\Omega-q^2\Gamma}\fr{\mu}{\bar{r}}-\fr{\Lambda}{3(\Omega-q^2\Gamma)}\bar{r}^2.
    \end{align}
Thus, the conformal/disformal transformation amounts to a redefinition of the coordinates and parameters.

It is now straightforward to construct the transformation law for perturbation variables:~One starts from the perturbed metric~$g_\mn=g^{(0)}_\mn+h_\mn$ in the original frame, performs the transformation~\eqref{disformal}, changes the coordinates from $(\tau,\rho)$ to $(\bar{\tau},\bar{\rho})$, and then the deviation from \eqref{disformal_SdS} can be identified as the perturbation variables in the new frame.
The result is summarized as follows:
    \be
    \begin{split}
    &\bar{h}_0=\fr{h_0}{\sqrt{\Omega-q^2\Gamma}}, \qquad
    \bar{h}_1=\fr{h_1}{\sqrt{\Omega-q^2\Gamma}}, \qquad
    \bar{h}_2=h_2, \\
    &\bar{H}_0=\fr{\Omega+q^2\Omega_X-q^4\Gamma_X}{\Omega-q^2\Gamma}H_0+\fr{2q(\Gamma+\Omega_X-q^2\Gamma_X)}{\Omega-q^2\Gamma}\dot{\delta\phi}, \qquad
    \bar{H}_1=\fr{\Omega}{\Omega-q^2\Gamma}H_1+\fr{q\Gamma}{\Omega-q^2\Gamma}\delta\phi', \\
    &\bar{H}_2=H_2-\fr{q^2\Omega_X}{\Omega}H_0-\fr{2q\Omega_X}{\Omega}\dot{\delta\phi}, \qquad
    \bar{\alpha}=\fr{\alpha}{\sqrt{\Omega-q^2\Gamma}}+\fr{q\Gamma}{\sqrt{\Omega-q^2\Gamma}}\fr{\delta\phi}{\Omega r^2}, \\
    &\bar{\beta}=\fr{\beta}{\sqrt{\Omega-q^2\Gamma}}, \qquad
    \bar{K}=K-\fr{q^2\Omega_X}{\Omega}H_0-\fr{2q\Omega_X}{\Omega}\dot{\delta\phi}, \qquad
    \bar{G}=G.
    \end{split}
    \ee
Hence, the conformal/disformal transformation does not mix the odd- and even-parity perturbations.

Having derived the transformation law for perturbation variables, one can derive the field redefinition~\eqref{map2Horndeski} for even-parity perturbations.
Recall that we chose the gauge fixing~$K=G=\delta\phi=0$ in \S\ref{ssec:multipole_even}.
Performing the conformal/disformal transformation, we have $\bar{K}=-(q^2\Omega_X/\Omega)H_0$ in the new frame, so we make a gauge transformation to set $\bar{K}\to 0$.
As a result, the perturbation variables for even modes in the new frame are
    \be
    \begin{split}
    &\bar{H}_0=\fr{\Omega+q^2\Omega_X-q^4\Gamma_X}{\Omega-q^2\Gamma}H_0, \qquad
    \bar{H}_1=\fr{\Omega}{\Omega-q^2\Gamma}\brb{H_1+\fr{q^2\Omega_X}{2\Omega}(1-A)\bra{\fr{r}{\sqrt{1-A}}H_0}^{\boldsymbol{\cdot}}\,}, \\
    &\bar{H}_2=H_2+\fr{q^2\Omega_X}{\Omega}\fr{r}{\sqrt{1-A}}H_0', \qquad
    \bar{\alpha}=\fr{\alpha}{\sqrt{\Omega-q^2\Gamma}}, \qquad
    \bar{\beta}=\fr{1}{\sqrt{\Omega-q^2\Gamma}}\bra{\beta+\fr{q^2\Omega_X}{2\Omega}\fr{\sqrt{1-A}}{r}H_0}, \label{even_disformal}
    \end{split}
    \ee
and $\bar{K}=\bar{G}=\delta\phi=0$.
From \eqref{even_disformal}, we see that $H_1$, $H_2$, and $\beta$ are mixed with $H_0$ and its derivatives.
It was shown in \cite{Achour:2016rkg} that class Ia DHOST theories described by the action~\eqref{qDHOST} can be mapped into the Horndeski subclass by the conformal/disformal transformation with
    \be
    \fr{\Omega_X}{\Omega}=\fr{4F_{2X}-2A_1+XA_3}{4(F_2-XA_1)}, \qquad
    \fr{\Gamma_X}{\Omega}=\fr{2A_1(A_1-2F_{2X})-A_3(2F_2-XA_1)}{4(F_2-XA_1)^2}.
    \ee
For this choice of $\Omega$ and $\Gamma$, we obtain
    \be
    \begin{split}
    \bar{H}_1&=\fr{\Omega}{\Omega-q^2\Gamma}\brb{H_1+\Up(1-A)\bra{\fr{r}{\sqrt{1-A}}H_0}^{\boldsymbol{\cdot}}\,}
    =\fr{\Omega}{\Omega-q^2\Gamma}\ti{H}_1, \\
    \bar{H}_2&=H_2+2\Up\fr{r}{\sqrt{1-A}}H_0'
    =\ti{H}_2, \\
    \bar{\beta}&=\fr{1}{\sqrt{\Omega-q^2\Gamma}}\bra{\beta+\Up\fr{\sqrt{1-A}}{r}H_0}
    =\fr{\ti{\beta}}{\sqrt{\Omega-q^2\Gamma}},
    \end{split}
    \ee
with $\Up$ defined in \eqref{Upsilon}, and thus arriving at the field redefinition~\eqref{map2Horndeski}.

\section{Hamiltonian analysis for even modes}\label{AppB}

In this appendix, we perform a Hamiltonian analysis for even modes with $\ell\ge 1$ to clarify the number of dynamical DOFs.
The analysis also includes special cases with any of $\Psi$, $\Phi$, and $\Pi$ vanishing.
We recall that the quadratic Lagrangian for even modes is given by
	\begin{align}
	\mL=&~
	j^2\ti{a}_4\bra{\dot{\ti{\beta}}-\alpha'}^2
	+\ti{b}_2\dot{\ti{H}}_2H_0+\ti{b}_3\dot{\ti{H}}_2\ti{H}_1+j^2\ti{b}_4\dot{\ti{H}}_2\alpha
	+j^2\ti{b}_5\dot{\ti{\beta}}\ti{H}_1 \nonumber \\
	&+\ti{b}_6H'_0\ti{H}_1+\ti{b}_7H'_0\ti{H}_2+j^2\ti{b}_8H'_0\ti{\beta}+j^2\ti{b}_9\alpha'\ti{H}_1 \nonumber \\
	&+\ti{c}_1H_0^2+j^2\ti{c}_3H_0\ti{H}_2+j^2\ti{c}_4H_0\alpha+j^2\ti{c}_5H_0\ti{\beta}+j^2\ti{c}_6\ti{H}_1^2+j^2\ti{c}_7\ti{H}_1\alpha+j^2\ti{c}_8\ti{H}_1\ti{\beta} \nonumber \\
	&+\ti{c}_9\ti{H}_2^2+j^2\ti{c}_{10}\ti{H}_2\alpha+j^2\ti{c}_{11}\ti{H}_2\ti{\beta}+j^2\ti{c}_{12}\alpha^2+j^2\ti{c}_{13}\ti{\beta}^2,
	\end{align}
where we have omitted an overall numerical factor [see \eqref{qlag2_even}].
The canonical momenta are computed as 
    \be
    \begin{split}
    &\pi_0\coloneqq \fr{\pa\mL}{\pa\dot{H}_0}=0, \qquad
    \pi_1\coloneqq \fr{\pa\mL}{\pa\dot{\ti{H}}_1}=0, \qquad
    \pi_2\coloneqq \fr{\pa\mL}{\pa\dot{\ti{H}}_2}=\ti{b}_2H_0+\ti{b}_3\ti{H}_1+j^2\ti{b}_4\alpha, \\
    &\pi_\alpha\coloneqq \fr{\pa\mL}{\pa\dot{\alpha}}=0, \qquad
    \pi_\beta\coloneqq \fr{\pa\mL}{\pa\dot{\ti{\beta}}}=2j^2\ti{a}_4\bra{\dot{\ti{\beta}}-\alpha'}+j^2\ti{b}_5\ti{H}_1, \qquad
    \end{split}
    \ee
and thus we have four primary constraints,
    \be
    \pi_0\approx 0, \qquad
    \pi_1\approx 0, \qquad
    \ti{\pi}_2\coloneqq \pi_2-\ti{b}_2H_0-\ti{b}_3\ti{H}_1-j^2\ti{b}_4\alpha\approx 0, \qquad
    \pi_\alpha\approx 0. \qquad
    \ee
The total Hamiltonian is given by 
    \be
    \begin{split}
    H_T&=\int d\rho \bra{\mH+v_0\pi_0+v_1\pi_1+v_2\ti{\pi}_2+v_\alpha\pi_\alpha}, \\
    \mH&\coloneqq \pi_2\dot{\ti{H}}_2+\pi_\beta\dot{\ti{\beta}}-\mL,
    \end{split}
    \ee
with $v_0$, $v_1$, $v_2$, and $v_\alpha$ being Lagrange multipliers.
The explicit form of $\mH$ is given by
    \begin{align}
    \mH= 
    &\fr{1}{4j^2\ti{a}_4}\bra{\pi_\beta-j^2\ti{b}_5\ti{H}_1}^2+\alpha'\pi_\beta \nonumber \\
    &-\Bigl(\ti{b}_6H'_0\ti{H}_1+\ti{b}_7H'_0\ti{H}_2+j^2\ti{b}_8H'_0\ti{\beta}+2j^2\ti{b}_5\alpha'\ti{H}_1+\ti{c}_1H_0^2+j^2\ti{c}_3H_0\ti{H}_2+j^2\ti{c}_4H_0\alpha+j^2\ti{c}_5H_0\ti{\beta} \nonumber \\
    &\qquad +j^2\ti{c}_6\ti{H}_1^2+j^2\ti{c}_7\ti{H}_1\alpha+j^2\ti{c}_8\ti{H}_1\ti{\beta}+\ti{c}_9\ti{H}_2^2+j^2\ti{c}_{10}\ti{H}_2\alpha+j^2\ti{c}_{11}\ti{H}_2\ti{\beta}+j^2\ti{c}_{12}\alpha^2+j^2\ti{c}_{13}\ti{\beta}^2\Bigr),
    \end{align}
where we have used $\ti{b}_9=\ti{b}_5$ [see \eqref{coeffs}].
It should be noted that $\ti{a}_4\propto F_2-XA_1\ne 0$, or otherwise gravitational waves do not propagate.
We also define $H\coloneqq \int d\rho\,\mH$ for convenience.
The consistency conditions for the primary constraints read
    \begin{align}
    \dot{\pi}_0&=\PB{\pi_0,H_T}\approx \PB{\pi_0,H}+\ti{b}_2v_2\approx 0, \label{ccp1} \\
    \dot{\pi}_1&=\PB{\pi_1,H_T}\approx \PB{\pi_1,H}+\ti{b}_3v_2\approx 0, \label{ccp2} \\
    \dot{\ti{\pi}}_2&=\PB{\ti{\pi}_2,H_T}\approx \PB{\ti{\pi}_2,H}-\ti{b}_2v_0-\ti{b}_3v_1-j^2\ti{b}_4v_\alpha\approx 0, \label{ccp3} \\
    \dot{\pi}_\alpha&=\PB{\pi_\alpha,H_T}\approx \PB{\pi_\alpha,H}+j^2\ti{b}_4v_2\approx 0. \label{ccp4}
    \end{align}
Since $\ti{b}_3\ne 0$, \eqref{ccp2} yields $v_2\approx -\ti{b}_3^{-1}\PB{\pi_1,H}$, which can be substituted into \eqref{ccp1} and \eqref{ccp4} to obtain the following secondary constraints:
    \begin{align}
    \chi_0&\coloneqq \PB{\pi_0,H}-\fr{\ti{b}_2}{\ti{b}_3}\PB{\pi_1,H}-\fr{\Phi}{F_2+q^2A_1}\ti{\pi}_2'\approx 0, \label{chi0}\\
    \chi_\alpha&\coloneqq \PB{\pi_\alpha,H}-\fr{j^2\ti{b}_4}{\ti{b}_3}\PB{\pi_1,H}+j^2r\bra{\fr{\ti{\pi}_2}{\sqrt{1-A}}}'\approx 0. \label{chia}
    \end{align}
Here, we have inserted the last terms in $\chi_0$ and $\chi_\alpha$, which are weakly vanishing, to cancel the terms with $H'_0,H'_1,\alpha'$ in $\chi_0$ and $\chi_\alpha$.
While the explicit forms of $\chi_0$ and $\chi_\alpha$ are still lengthy, thanks to the introduction of the last terms, their dependency on $H_0,H_1,\alpha$ is simple:
    \be
    \chi_0= 2 \Psi r^2 \sqrt{1-A}\,H_0 + \cdots, \qquad
    \chi_\alpha=-j^2(j^2-2)\ti{b}_4\alpha + \cdots,
    \ee
where the ellipses denote terms without $H_0$, $H_1$, or $\alpha$.
These processes significantly simplify the consistency conditions for the secondary constraints, which read
    \begin{align}
    \dot{\chi}_0&=\PB{\chi_0,H_T}\approx 
    \PB{\chi_0,H}+2\Psi r^2\sqrt{1-A}\,v_0\approx 0, \label{ccs1} \\
    \dot{\chi}_\alpha&=\PB{\chi_\alpha,H_T}\approx 
    \PB{\chi_\alpha,H}-j^2(j^2-2)\ti{b}_4v_\alpha\approx 0. \label{ccs2}
    \end{align}
Now, we have three equations~\eqref{ccp3}, \eqref{ccs1}, and \eqref{ccs2} for the remaining three Lagrange multipliers~$v_0$, $v_1$, and $v_\alpha$.
The determinant of the system of equations is proportional to $\Delta \coloneqq j^2(j^2-2)\Psi r^2\sqrt{1-A}\,\ti{b}_3\ti{b}_4$. 
Hence, so long as $j^2\ne 2$ (i.e., $\ell\ne 1$) and $\Psi\ne 0$, one can fix all the Lagrange multipliers. 
Also, under these conditions, we find that all the six constraints specified so far are of second class since the determinant of 
the matrix of Poisson brackets is proportional to $\Delta^2\ne 0$.
As a result, the number of physical DOFs of the system is given by
    \begin{align}
    &\fr{1}{2}\brb{\bra{\text{$\#$ of phase-space variables}}
    -2\times \bra{\text{$\#$ of first-class constraints}}
    -\bra{\text{$\#$ of second-class constraints}}} \nonumber \\
    &\qquad =\fr{1}{2}\bra{10-2\times 0-6}=2,
    \end{align}
thus arriving at the desired number of DOFs for even modes in single-field scalar-tensor theories.

It is also straightforward to count the number of DOFs for cases with $\ell=1$ and/or $\Psi=0$.
Here, we just summarize the results.

\begin{itemize}
\item 
First, let us consider the case where $\ell=1$ and $\Psi\ne 0$. In this case, \eqref{ccs2} yields a tertiary constraint and it reduces the number of DOFs to one, which is consistent with the result of \S\ref{ssec:dipole_even}.

\item
For the case where $\Psi=0$, we first assume $\Phi\ne 0$ or $\Pi\ne 0$. 
In this case, for $\ell\ge 2$ modes, \eqref{ccs1} yields a tertiary constraint, and as a result there is only one physical DOF.
For $\ell=1$, both \eqref{ccs1} and \eqref{ccs2} yield tertiary constraints, and hence no dynamical DOF is left.
As mentioned in \S\ref{ssec:multipole_even}, this result is reasonable as the class of theories with $\Psi(X)=0$ includes those having no scalar DOF, like the extended cuscutons~\cite{Iyonaga:2018vnu}.

\item 
Finally, let us consider the case where $\Psi=\Phi=\Pi=0$ at least at $X=-q^2$. 
In this case, the secondary constraint~$\chi_0\approx 0$ becomes trivial, which comes from the fact that the Lagrangian is independent of $H_0$ in this particular case.
As a result, the number of DOFs is two for $\ell\ge 2$, while the modes with $\ell=1$ have one DOF.\footnote{Given that $\Psi(X)=0$ would imply the absence of the scalar DOF, this result might seem counterintuitive. 
To address this issue, let us consider the extended cuscutons as a general class of theories with $\Psi(X)=0$.
If $\Phi(X)=\Pi(X)=0$ is imposed on \eqref{cuscuton}, we obtain $F_2-XA_1=0$, meaning the absence of gravitational waves.
Then, we have $\ti{a}_4=\ti{b}_3=0$ and the above discussion can no longer be applied.}
Interestingly, this case includes the case where the action has a local conformal symmetry, like mimetic gravity models~\cite{Chamseddine:2013kea,Sebastiani:2016ras,Takahashi:2017pje,Langlois:2018jdg}.
Indeed, the conformally invariant subclass of \eqref{qDHOST} has the coefficient functions of the form
    \be
    F_0\propto X^2, \qquad
    F_2\propto X, \qquad
    A_1=-\fr{X}{2}A_3={\rm const},
    \ee
which satisfies $\Psi(X)=\Phi(X)=\Pi(X)=0$ at the functional level.
This fact suggests that the quadratic action for perturbations enjoys a gauge symmetry associated with the conformal invariance, which explains why $H_0$ disappears from the action.
It should be noted that the mimetic gravity models are known to have ghost/gradient instabilities on a cosmological background~\cite{Takahashi:2017pje,Langlois:2018jdg}.
Hence, we expect that a similar instability would show up in the present case.
\end{itemize}

\section{Characteristic analysis}\label{AppC}

As we pointed out earlier, the analysis in \S\ref{ssec:monopole} does not apply if any of $\Psi$, $\Phi$, and $\Pi$ vanishes.
In this appendix, we perform a characteristic analysis, which applies to such pathological cases.
In what follows, we study a system defined in $1+1$ dimensions.
First, we describe our strategy for a general system, and then consider its application to toy models in Appendices~\ref{AppC1} and \ref{AppC2} and the case of monopole perturbations in Appendix~\ref{AppC3}.

Let us use $\tau$ and $\rho$ as the coordinate variables as in the main text and assume a diagonal metric with Lorentzian signature.
We follow an algorithm based on the ``Kronecker canonical form'' (KCF) of a matrix pencil, which applies to a set of first-order linear partial differential-algebraic equations in $1+1$ dimensions (see Appendices~A and B\,1 of \cite{Motloch:2016msa} for details).
In this algorithm, we first derive the EOMs from a given Lagrangian in $1+1$ dimensions and recast them into the first-order form,
    \be
    \bA\dot{\vu}+\bB\vu'+\bC\vu=0, \label{PDAE}
    \ee
where $\bA$, $\bB$, $\bC$ are $M\times N$ coefficient matrices and $\vu$ is an $N$-dimensional vector of dependent variables.
In the characteristic analysis, we focus on how the highest derivatives appear in the EOMs.
For this purpose, it is useful to construct a matrix pencil~$\bA+\la \bB$, where $\la$ is a label to keep track the spatial derivative.
The following transformation amounts to a field redefinition and a linear combination of the EOMs:
\be \tilde \vu= \bQ^{-1}\vu, \qquad 
\tilde \bA = \bP\bA\bQ, \qquad 
\tilde \bB = \bP\bB\bQ, \qquad 
\tilde \bC = \bP\bC\bQ + \bP\bA\dot \bQ + \bP\bB\bQ',  \ee
where $\bP$ and $\bQ$ are invertible matrices having the size of $M\times M$ and $N\times N$, respectively.
As a result, $\bA+\la \bB$ transforms as 
\be \bA+\lambda \bB \to \tilde \bA + \lambda \tilde \bB = \bP(\bA+\lambda \bB)\bQ.  \ee
A suitable choice of $\bP$ and $\bQ$ puts $\ti{\bA}+\la\ti{\bB}$ into its KCF~\cite{Berger:2012}, i.e., 
    \be
    \ti{\bA}+\la \ti{\bB}={\rm diag}\bra{\bL_{m_1},\cdots,\bL_{m_u},\bL_{n_1}^P,\cdots,\bL_{n_o}^P,\bR_{k_1}(\ka_1),\cdots,\bR_{k_r}(\ka_r)}.
    \ee
Here, $\bL_m$ is an $m\times(m+1)$ bidiagonal matrix pencil and $\bL_m^P$ denotes its pertransposition defined by
    \be
    \bL_m=\bem
    0 & 1 &  &  \\
     & \ddots & \ddots &  \\
     &  & 0 & 1
    \eem+\la\bem
    1 & 0 &  &  \\
     & \ddots & \ddots &  \\
     &  & 1 & 0
    \eem, \qquad
    \bL_m^P=\bem
    1 & & \\
    0 & \ddots & \\
     & \ddots & 1 \\
     & & 0
    \eem+\la\bem
    0 & & \\
    1 & \ddots & \\
     & \ddots & 0 \\
     & & 1
    \eem,
    \ee
and $\bR_k(\ka)$ is a $k\times k$ regular matrix pencil of the following form:
    \be
    \bR_k(\ka)=\left\{
    \begin{array}{ll}
    \la \bI_k-\bem
    \ka & & & \\
    1 & \ka & & \\
     & \ddots & \ddots & \\
     & & 1 & \ka
    \eem & \text{for } |\ka|<\infty, \\
    \la \bem
    0 & & & \\
    1 & 0 & & \\
     & \ddots & \ddots & \\
     & & 1 & 0
    \eem-\bI_k & \text{for } |\ka|=\infty,
    \end{array}\right. 
    \ee
with $\bI_k$ being a $k\times k$ identity matrix.
Then, one can obtain characteristics from the KCF. 
Before applying this procedure to the Lagrangian for monopole perturbations, we study some toy examples to demonstrate how the above procedure works and what the KCF tells us about the physics of a given system.

\subsection{Free point particle}\label{AppC1}

As a simplest example, let us consider the following Lagrangian for a free particle in $1+1$ dimensions:
\be \mL=\fr{a}{2}\dot\phi^2 - \fr{b}{2}\phi'^2, \label{toy1} \ee
where $a,b$ are nonnegative constants.
The EOM for $\phi$ is given by
\be a\ddot\phi-b\phi''=0. \ee
Provided that $a>0$ and $b>0$, one can read off the squared sound speed from this EOM as
    \be
    c_s^2=\fr{g_{\rho\rho}}{|g_{\tau\tau}|}\fr{b}{a}.
    \ee
To reduce the system into the first-order form, let us introduce auxiliary fields,
\be \phi_\tau=\dot\phi, \qquad \phi_\rho=\phi' .  \ee
Note that they satisfy 
\be \phi_\tau'=\dot \phi_\rho . \ee
We then sort out the four equations as 
    \be
    \dot{\phi}-\phi_\tau=0, \qquad
    \phi'-\phi_\rho=0, \qquad
    a\dot{\phi}_\tau-b\phi_\rho'=0, \qquad
    \phi_\tau'-\dot{\phi}_\rho=0.
    \ee
Denoting $\vu=(\phi,\phi_\tau ,\phi_\rho )^T$, this system is written in the form~\eqref{PDAE} with
\be 
\bA= \bem 1 & 0 & 0 \\
0 & 0 & 0 \\
0 & a & 0 \\
0 & 0 & 1
\eem , \qquad
\bB= \bem 0 & 0 & 0 \\
1 & 0 & 0 \\
0 & 0 & -b \\
0 & -1 & 0
\eem , \qquad
\bC= \bem 0 & -1 & 0 \\
0 & 0 & -1 \\
0 & 0 & 0 \\
0 & 0 & 0
\eem ,
\ee
from which we construct the following matrix pencil:
\be 
\bA+\lambda \bB = \bem 1 & 0 & 0 \\
\lambda & 0 & 0 \\
0 & a & -b\lambda \\
0 & -\lambda & 1
\eem .
\ee

For $a>0$ and $b>0$, the KCF of the matrix pencil is given by 
\be 
\tilde \bA+\lambda \tilde \bB = \bem 1 & 0 & 0 \\
\lambda & 0 & 0 \\
0 & \displaystyle\lambda-\sqrt{\fr{a}{b}} & 0 \\
0 & 0 & \displaystyle\lambda+\sqrt{\fr{a}{b}}
\eem
={\rm diag}\bra{ \bL_1^P, \bR_1\bra{\sqrt{\fr{a}{b}}}, \bR_1\bra{-\sqrt{\fr{a}{b}}} },
\ee
where $\tilde \bA+\lambda \tilde \bB = \bP (\bA+\lambda \bB)\bQ$ with 
\be \label{KF1} \bP= \bem 1 & 0 & 0 & 0 \\
0 & 1 & 0 & 0 \\
0 & 0 & \displaystyle-\fr{1}{2b} & \displaystyle-\fr{1}{2}\sqrt{\fr{a}{b}} \\
0 & 0 & \displaystyle-\fr{1}{2b} & \displaystyle\fr{1}{2}\sqrt{\fr{a}{b}}
\eem
,\qquad 
\bQ= \bem 1 & 0 & 0  \\
0 & \displaystyle\sqrt{\fr{b}{a}} & \displaystyle-\sqrt{\fr{b}{a}} \\
0 & 1 & 1
\eem .
\ee

For $a>0$ and $b=0$, the EOM for $\phi$ reduces to
\be \ddot\phi=0, \ee
which amounts to $c_s^2=0$.
The first-order system then takes the form 
    \be
    \dot{\phi}-\phi_\tau=0, \qquad
    \dot{\phi}_\tau=0,
    \ee
which corresponds to the following matrix pencil:
\be 
\bA+\lambda \bB = \bem 1 & 0 \\
0 & 1 
\eem .
\ee
The KCF is given by 
\be \label{KF2}
\tilde \bA+\lambda \tilde \bB = \bem -1 & 0 \\
0 & -1 
\eem
={\rm diag}\bra{\bR_1(\infty),\bR_1(\infty)}.
\ee

For $a= 0$ and $b>0$, the EOM for $\phi$ reduces to
\be \phi''=0, \ee
which amounts to $c_s^2=\infty$.
The first-order system then takes the form 
    \be
    \phi'-\phi_\rho=0, \qquad
    \phi'_\rho=0.
    \ee
This system of equations corresponds to the following matrix pencil:
    \be 
    \bA+\lambda \bB = \bem \lambda & 0 \\
    0 & \lambda 
    \eem 
    ={\rm diag}\bra{\bR_1(0),\bR_1(0)}, \label{KF3}
    \ee
which is itself of the KCF.

The above toy example shows a link between the sound speed and the regular part of the KCF.
Given the KCF of the matrix pencil of a system of first-order equations, we can define the sound speed if we view the system as a second-order one.
For the regular part of the KCF given by
    \be
    {\rm diag}\bra{\bR_1(\ka),\bR_1(-\ka)},
    \ee
we define the sound speed as
    \be 
    c_s^2=\fr{g_{\rho\rho}}{|g_{\tau\tau}|}\ka^{-2}. \label{sound_speed}
    \ee
For $\ka=0$ (or $\ka=\infty$), the sound speed is divergent (or vanishing), suggesting the absence of the kinetic term (or gradient term, respectively) in the corresponding second-order system.
Indeed, from the KCFs~\eqref{KF1}, \eqref{KF2}, \eqref{KF3}, we can read off the squared sound speed as $c_s^2=\fr{b}{a}, 0, \infty$, respectively, which is consistent with the one obtained from the second-order EOM for $\phi$.
We can generalize this definition for systems with multiple DOFs, for which we typically have the KCF with a regular part of the following form:
    \be
    {\rm diag}\bra{\bR_1(\ka_1),\bR_1(-\ka_1),\bR_1(\ka_2),\bR_1(-\ka_2),\cdots}.
    \ee
In this case, the sound speed for each mode is defined as
    \be 
    c_{s1}^2=\fr{g_{\rho\rho}}{|g_{\tau\tau}|}\ka_1^{-2}, \qquad c_{s2}^2=\fr{g_{\rho\rho}}{|g_{\tau\tau}|}\ka_2^{-2}, \qquad \cdots . 
    \ee
Also, even if a system cannot be identified as an ordinary second-order system, it is natural to generalize the above definition of the sound speed.
For instance, for the KCF whose regular part is of the form~$\bR_2(\ka)$, we define $c_s^2$ as in \eqref{sound_speed}.
Likewise, one can read off the sound speed from the regular part of the KCF associated with the EOMs of a given system.
Such a definition of the sound speed is especially useful for involved systems since the KCF can be obtained completely in an algorithmic manner.

A caveat is that the implication of the absence of the kinetic or gradient term depends on theories.
If one considers a system with time dependence only, the gradient term is intrinsically absent, and hence it is not pathology.
Specific examples are analytic mechanics of interacting point particles and cosmological evolution of a homogeneous scalar field on a homogeneous and isotropic background.
Likewise, if one focuses on the evolution of a static scalar field on a spherically symmetric background, then the kinetic term is intrinsically absent, which is also no problem.
On the other hand, if one considers the evolution of some fields in $1+1$ dimensions in an EFT, the diverging/vanishing sound speed, corresponding respectively to the absence of the kinetic/gradient term, would imply the strong coupling. 
In such a case, the characteristic analysis serves as an efficient algorithm for diagnostics.

Another caveat is that the KCF does not necessarily contain a regular part.
Such a case corresponds to the absence of dynamical DOFs, as illustrated by the toy example in the next section.

\subsection{Toy model with no dynamical DOF}\label{AppC2}

Next, let us consider the following example:
    \be
    \mL=\psi\bra{\phi'+\dot{\chi}}-\fr{1}{2}\psi^2, \label{toy2}
    \ee
which we shall see is related to some special case of the monopole analysis below.
Note that $\chi$ and $\psi$ here have nothing to do with those appeared in the main text.

First, let us perform a Hamiltonian analysis to show that the system described by \eqref{toy2} has no dynamical DOF.
The canonical momenta are computed as
    \be
    \pi_\phi\coloneqq \fr{\pa\mL}{\pa\dot{\phi}}=0, \qquad
    \pi_\psi\coloneqq \fr{\pa\mL}{\pa\dot{\psi}}=0, \qquad
    \pi_\chi\coloneqq \fr{\pa\mL}{\pa\dot{\chi}}=\psi, \qquad
    \ee
and thus we have three primary constraints,
    \be
    \pi_\phi\approx 0, \qquad
    \pi_\psi\approx 0, \qquad
    \ti{\pi}_\chi\coloneqq \pi_\chi-\psi\approx 0.
    \ee
The total Hamiltonian is given by
    \be
    \begin{split}
    H_T&=\int d\rho \bra{\mH+u_\phi\pi_\phi+u_\psi\pi_\psi+u_\chi\ti{\pi}_\chi}, \\
    \mH&\coloneqq \pi_\chi\dot{\chi}-\mL=-\psi\phi'+\fr{1}{2}\psi^2,
    \end{split}
    \ee
where $u_\phi$, $u_\psi$, and $u_\chi$ are Lagrange multipliers.
The consistency conditions for $\pi_\psi\approx 0$ and $\ti{\pi}_\chi\approx 0$ respectively fix the Lagrange multipliers as $u_\chi\approx \psi-\phi'$ and $u_\psi\approx 0$, while the consistency condition for $\pi_\phi\approx 0$ yields the following secondary constraint:
    \be
    \dot{\pi}_\phi=\PB{\pi_\phi,H_T}\approx -\psi'\approx 0.
    \ee
One can show that the consistency condition for this secondary constraint is automatically satisfied and the Lagrange multiplier~$u_\phi$ remains unfixed.
As a result, we end up with two first-class constraints~$\pi_\phi\approx 0$, $\psi'\approx 0$ and two second-class constraints~$\pi_\psi\approx 0$, $\ti{\pi}_\chi\approx 0$.
Then, the number of DOFs of the system can be calculated as
    \be
    \fr{1}{2}\bra{6-2\times 2-2}=0.
    \ee

Now, let us perform the characteristic analysis for the model~\eqref{toy2}.
The EOMs are given by
    \be
    \phi'+\dot{\chi}-\psi=0, \qquad
    \dot{\psi}=0, \qquad
    \psi'=0.
    \ee
From these EOMs, one can construct the matrix pencil~$\bA+\la\bB$ as
    \be
    \bA+\la\bB=\bem
    \la & 1 & 0 \\
    0 & 0 & 1 \\
    0 & 0 & \la
    \eem
    ={\rm diag}\bra{\bL_1,\bL_1^P}, \label{toy2KCF}
    \ee
which is itself of the KCF.
Note that there is no regular part in the KCF.
Combined with the result of the Hamiltonian analysis, it is natural to connect such a form of KCF with the absence of dynamical DOFs.

\subsection{Application to monopole perturbations}\label{AppC3}

Now we apply the above procedure to the monopole perturbation.
Instead of \eqref{qlag3_mono}, we start from the following Lagrangian, which is obtained by taking $\ell=0$ in \eqref{qlag2_even}:
    \be
    \mL=
	\ti{b}_2\dot{\ti{H}}_2H_0+\ti{b}_3\dot{\ti{H}}_2\ti{H}_1
	+\ti{b}_6H'_0\ti{H}_1+\ti{b}_7H'_0\ti{H}_2
	+\ti{c}_1H_0^2+\ti{c}_9\ti{H}_2^2, \label{qlag2_mono}
    \ee
where we have omitted an overall numerical factor.
The EOMs read
    \be
    \begin{split}
    \mE_0&\coloneqq \fr{\pa\mL}{\pa H_0}-\bra{\fr{\pa\mL}{\pa H_0'}}'=\ti{b}_2\dot{\ti{H}}_2-\ti{b}_6\ti{H}_1'-\ti{b}_7\ti{H}_2'+2\ti{c}_1H_0-\ti{b}_6'\ti{H}_1-\ti{b}_7'\ti{H}_2=0, \\
    \mE_1&\coloneqq \fr{\pa\mL}{\pa \ti{H}_1}=\ti{b}_3\dot{\ti{H}}_2+\ti{b}_6H_0'=0, \\
    \mE_2&\coloneqq \fr{\pa\mL}{\pa \ti{H}_2}-\bra{\fr{\pa\mL}{\pa \dot{\ti{H}}_2}}^{\boldsymbol{\cdot}}=-\ti{b}_2\dot{H}_0-\ti{b}_3\dot{\ti{H}}_1+\ti{b}_7H_0'-\dot{\ti{b}}_2H_0-\dot{\ti{b}}_3\ti{H}_1+2\ti{c}_9\ti{H}_2=0.
    \end{split}
    \ee
Also, one can construct a linearly independent first-order equation as
    \be
    \mE_3\coloneqq \ti{b}_3\dot{\mE}_0-\ti{b}_2\dot{\mE}_1+\ti{b}_7\mE_1'-\ti{b}_6\mE_2'
    =g_1\dot{H}_0+g_2\dot{\ti{H}}_2+g_3H_0'+g_4\ti{H}_2'+g_5H_0+g_6\ti{H}_2=0,
    \ee
where
    \be
    \begin{split}
    &g_1\coloneqq 2\ti{b}_3\ti{c}_1+\ti{b}_6\ti{b}_2', \qquad
    g_2\coloneqq \ti{b}_3\dot{\ti{b}}_2-\ti{b}_2\dot{\ti{b}}_3, \qquad
    g_3\coloneqq \ti{b}_6\dot{\ti{b}}_2-\ti{b}_2\dot{\ti{b}}_6, \\
    &g_4\coloneqq -\bra{2\ti{b}_6\ti{c}_9+\ti{b}_3\dot{\ti{b}}_7}, \qquad
    g_5\coloneqq 2\ti{b}_3\dot{\ti{c}}_1+\ti{b}_6\dot{\ti{b}}_2', \qquad
    g_6\coloneqq -\bra{2\ti{b}_6\ti{c}_9'+\ti{b}_3\dot{\ti{b}}_7'}.
    \end{split}
    \ee
By use of \eqref{coeffs}, the first four coefficients can be written explicitly as
    \be
    \begin{split}
    &g_1=8(F_2+q^2A_1)\Psi r^3\sqrt{1-A}, \qquad
    g_2=8(F_2+q^2A_1)\Phi r^2A', \\
    &g_3=8\Phi^2r^2A'=\fr{\ti{b}_6}{\ti{b}_3}g_2, \qquad
    g_4=8q^4\Pi r\sqrt{1-A}.
    \end{split}
    \ee
Denoting $\vu=(H_0,\ti{H}_1,\ti{H}_2)^T$, this system is written in the form~\eqref{PDAE} with
    \be
    \bA=\bem
    0 & 0 & \ti{b}_2 \\
    0 & 0 & \ti{b}_3 \\
    -\ti{b}_2 & -\ti{b}_3 & 0 \\
    g_1 & 0 & g_2
    \eem, \qquad
    \bB=\bem
    0 & -\ti{b}_6 & -\ti{b}_7 \\
    \ti{b}_6 & 0 & 0 \\
    \ti{b}_7 & 0 & 0 \\
    g_3 & 0 & g_4
    \eem, \qquad
    \bC=\bem 
    2\ti{c}_1 & -\ti{b}_6' & -\ti{b}_7' \\
    0 & 0 & 0 \\
    -\dot{\ti{b}}_2 & -\dot{\ti{b}}_3 & 2\ti{c}_9 \\
    g_5 & 0 & g_6
    \eem,
    \ee
from which we construct the following matrix pencil:
    \be
    \bA+\la\bB=
    \bem
    0 & -\ti{b}_6\la & \ti{b}_2-\ti{b}_7\la \\
    \ti{b}_6\la & 0 & \ti{b}_3 \\
    -\ti{b}_2+\ti{b}_7\la & -\ti{b}_3 & 0 \\
    g_1+g_3\la & 0 & g_2+g_4\la
    \eem.
    \ee
For $\Psi\Phi\Pi\ne 0$, the KCF is given by
\be 
\tilde \bA+\lambda \tilde \bB = \bem 1 & 0 & 0 \\
\lambda & 0 & 0 \\
0 & \lambda-\ka_\ast & 0 \\
0 & 0 & \lambda+\ka_\ast
\eem
={\rm diag}\bra{ \bL_1^P, \bR_1(\ka_\ast), \bR_1(-\ka_\ast) },\qquad
\ka_\ast\coloneqq \sqrt{\fr{g_1g_2}{g_3g_4}}, \label{KCF_mono1}
\ee
where the transformation matrices~$\bP$ and $\bQ$ are
\be \bP= \bem 0 & \displaystyle\fr{\ti{b}_7}{\ti{b}_3\ti{b}_6} & \displaystyle-\fr{1}{\ti{b}_3} & 0 \\
\displaystyle-\fr{1}{\ti{b}_6} & \displaystyle\fr{\ti{b}_2}{\ti{b}_3\ti{b}_6} & 0 & 0 \\
0 & \displaystyle\fr{g_4\ka_\ast+g_2}{2\ti{b}_3g_4} & 0 & \displaystyle-\fr{1}{2g_4} \\
0 & \displaystyle\fr{g_4\ka_\ast-g_2}{2\ti{b}_3g_4} & 0 & \displaystyle\fr{1}{2g_4}
\eem
,\qquad 
\bQ= \bem 0 & \displaystyle\fr{g_4\ka_\ast}{g_1} & \displaystyle\fr{g_4\ka_\ast}{g_1} \\
1 & \displaystyle\fr{g_4\ka_\ast}{\ti{b}_3g_1}(\ti{b}_7\ka_\ast-\ti{b}_2) & \displaystyle-\fr{g_4\ka_\ast}{\ti{b}_3g_1}(\ti{b}_7\ka_\ast+\ti{b}_2) \\
0 & -1 & 1
\eem .
\ee
From the KCF~\eqref{KCF_mono1}, one can read off the squared sound speed as
    \be
    c_s^2=\fr{g_{\rho\rho}}{|g_{\tau\tau}|}\ka_\ast^{-2}=(1-A)\fr{g_1g_2}{g_3g_4}
    =\fr{q^4\Phi \Pi}{\Psi (F_2+q^2A_1)^2}\fr{1-A}{r^2},
    \ee
which is nothing but the squared sound speed~\eqref{ss_mono} for monopole perturbations.

As we pointed out earlier, the derivation of \eqref{ss_mono} does not apply if any of $\Psi$, $\Phi$, and $\Pi$ vanishes.
However, one can obtain the KCF and the characteristic analysis still works for such pathological cases.
Before performing a case-by-case analysis, we summarize the results in the \hyperref[table]{Table} below.
\vskip-\baselineskip
\begin{table}[h]
\renewcommand\thetable{\!\!}
\newcolumntype{C}[1]{>{\hfil}m{#1}<{\hfil}}
\centering
\caption{The squared sound speed for each case.}
\begin{tabular}{C{15mm}C{25mm}C{30mm}C{25mm}} \hline\hline
     & $\Phi\Pi\ne 0$ & $\Phi=0$ or $\Pi=0$ & $\Phi=\Pi=0$ \\ \hline
    $\Psi\ne 0$ & $c_s^2$ finite & $c_s^2=0$ & $c_s^2=0$ \\
    $\Psi=0$ & $c_s^2=\infty$ & $c_s^2=\infty$ & $c_s^2=0$ \\ \hline\hline
\end{tabular}\label{table}
\end{table}

\noindent
For $\Psi\ne 0$, the sound speed vanishes if $\Phi=0$ and/or $\Pi=0$.
On the other hand, for $\Psi=0$, we have $c_s^2=0$ if $\Phi=\Pi=0$, or otherwise the monopole perturbation is nondynamical.
In what follows, we study each case and show the corresponding KCFs.

\begin{itemize}
\item
For $\Pi=0$ (i.e., $g_4=g_6=0$) and $\Psi\Phi\ne 0$, the KCF is given by
\be 
\tilde \bA+\lambda \tilde \bB = \bem 1 & 0 & 0 \\
\lambda & 0 & 0 \\
0 & -1 & 0 \\
0 & \la & -1
\eem
={\rm diag}\bra{ \bL_1^P, \bR_2(\infty) }. \label{Pi=0}
\ee
As mentioned in Appendix~\ref{AppC1}, for this KCF, it is natural to assign $c_s^2=0$.
This case includes the simplest case of the k-essence theory, which amounts to
    \be
    F_{0XX}\ne 0, \qquad
    F_2={\rm const}\ne 0, \qquad
    A_1=0, \qquad
    A_3=0,
    \ee
in \eqref{qDHOST}.

\item
For $\Psi=0$ (i.e., $g_1=g_5=0$) and $\Phi\Pi\ne 0$, the KCF is given by
\be 
\tilde \bA+\lambda \tilde \bB = \bem 1 & 0 & 0 \\
\lambda & 0 & 0 \\
0 & \la & 0 \\
0 & -1 & \la
\eem
={\rm diag}\bra{ \bL_1^P, \bR_2(0) }.
\ee
For this KCF, we assign $c_s^2=\infty$, meaning that there is no physical DOF.
This result is reasonable since, as we pointed out earlier, this case includes theories where the scalar DOF does not propagate~\cite{Lin:2017oow,Chagoya:2018yna,Aoki:2018zcv,Afshordi:2006ad,Iyonaga:2018vnu,Iyonaga:2020bmm,Gao:2019twq}.

\item For $\Phi=0$ (i.e., $\ti{b}_2=\ti{b}_6=g_2=g_3=0$) and $\Psi\Pi\ne 0$, the KCF is given by
\be 
\tilde \bA+\lambda \tilde \bB = \bem 1 & 0 & 0 \\
\lambda & 0 & 0 \\
0 & -1 & 0 \\
0 & \la & -1
\eem
={\rm diag}\bra{ \bL_1^P, \bR_2(\infty) }.
\ee
As in the case of \eqref{Pi=0}, we have $c_s^2=0$ for this KCF.

\item For $\Psi=\Phi=0$ (i.e., $\ti{b}_2=\ti{b}_6=\ti{c}_1=g_1=g_2=g_3=g_5=0$) and $\Pi\ne 0$, we obtain a constraint,
    \be
    \ti{H}_2=\fr{g_4\mE_0+\ti{b}_7\mE_3}{\ti{b}_7g_6-\ti{b}_7'g_4}=0, 
    \ee
which allows us to eliminate $\ti{H}_2$ from the system.
As a result, we are left with a single equation,
    \be
    \mE_2=-\ti{b}_3\dot{\ti{H}}_1+\ti{b}_7H_0'-\dot{\ti{b}}_3\ti{H}_1=0.
    \ee
Correspondingly, we take $\vu=(H_0,\ti{H}_1)^T$ and construct a smaller matrix pencil~$\bA+\la\bB$.
The KCF is given by
    \be
    \ti{\bA}+\la\ti{\bB}=\bem
    \la & 1
    \eem
    =\bL_1.
    \ee
As was the toy model~\eqref{toy2}, there is no regular part in the KCF, and thus no dynamical DOFs.

\item For $\Psi=\Pi=0$ (i.e., $g_1=g_4=g_5=g_6=0$) and $\Phi\ne 0$, one finds that $\mE_3=0$ is no longer an independent equation.
From the remaining equations~$\mE_0=\mE_1=\mE_2=0$, one can construct the matrix pencil~$\bA+\la\bB$, which has the following KCF:
\be 
\tilde \bA+\lambda \tilde \bB = \bem \la & 1 & 0 \\
0 & 0 & 1 \\
0 & 0 & \la
\eem
={\rm diag}\bra{ \bL_1, \bL_1^P }.
\ee
This KCF is the same as \eqref{toy2KCF} for the toy model~\eqref{toy2}, which implies the absence of dynamical DOFs.

\item For $\Phi=\Pi=0$ (i.e., $\ti{b}_2=\ti{b}_6=\ti{b}_7=g_2=g_3=g_4=g_6=0$) and $\Psi\ne 0$, we obtain a constraint,
    \be
    H_0=\fr{1}{2\ti{c}_1}\mE_0=0,
    \ee
by which we remove $H_0$ from the system.
Then, we are left with the following two equations:
    \be
    \mE_1=\ti{b}_3\dot{\ti{H}}_2=0, \qquad
    \mE_2=-\ti{b}_3\dot{\ti{H}}_1-\dot{\ti{b}}_3\ti{H}_1+2\ti{c}_9\ti{H}_2=0.
    \ee
We now take $\vu=(\ti{H}_1,\ti{H}_2)^T$ and construct the matrix pencil~$\bA+\la\bB$ from the above equations.
The KCF is given by
\be
\tilde \bA+\lambda \tilde \bB = \bem -1 & 0 \\
0 & -1 
\eem
={\rm diag}\bra{\bR_1(\infty),\bR_1(\infty)},
\ee
corresponding to $c_s^2=0$.

\item For $\Psi=\Phi=\Pi=0$ (i.e., $\ti{b}_2=\ti{b}_6=\ti{b}_7=\ti{c}_1=g_1=g_2=g_3=g_4=g_5=g_6=0$), the equations~$\mE_0=\mE_3=0$ are trivially satisfied and we are left with the same set of equations as in the previous case.
Therefore, we have one dynamical DOF with $c_s^2=0$. 
\end{itemize}


\bibliographystyle{mybibstyle}
\bibliography{bib}

\end{document}